\documentclass[fleqn,usenatbib]{mnras}

\usepackage{newtxtext,newtxmath}

\usepackage[T1]{fontenc}

\usepackage{xcolor}

\def\magenta\magenta{\color{magenta}}

\newbox\grsign \setbox\grsign=\hbox{$>$} \newdimen\grdimen \grdimen=\ht\grsign
\newbox\simlessbox \newbox\simgreatbox
\setbox\simgreatbox=\hbox{\raise.5ex\hbox{$>$}\llap
     {\lower.5ex\hbox{$\sim$}}}\ht1=\grdimen\dp1=0pt
\setbox\simlessbox=\hbox{\raise.5ex\hbox{$<$}\llap
     {\lower.5ex\hbox{$\sim$}}}\ht2=\grdimen\dp2=0pt

\def\simless{\mathrel{\copy\simlessbox}}

\newbox\simppropto
\setbox\simppropto=\hbox{\raise.5ex\hbox{$\sim$}\llap
     {\lower.5ex\hbox{$\propto$}}}\ht2=\grdimen\dp2=0pt

\DeclareRobustCommand{\VAN}[3]{#2}
\let\VANthebibliography\thebibliography
\def\thebibliography{\DeclareRobustCommand{\VAN}[3]{##3}\VANthebibliography}

\usepackage{graphicx}	
\usepackage{amsmath}	

\title[Modelling the proto-Galaxy]{Modelling the density and mass of the Milky Way's proto-galaxy components with \textsl{APOGEE-Gaia}}

\author[D. Horta $\&$ R.P. Schiavon]{
Danny Horta$^{1}$\thanks{E-mail: dhortadarrington@gmail.com}
and Ricardo P. Schiavon$^{2}$
\\
$^{1}$Center for Computational Astrophysics, Flatiron Institute, 162 Fifth Ave, New York, NY 10010, USA\\
$^{2}$Astrophysics Research Institute, 146 Brownlow Hill, Liverpool, L3 5RF, UK
}

\date{Accepted XXX. Received YYY; in original form ZZZ}

\pubyear{2024}

\begin{document}
\label{firstpage}
\pagerange{\pageref{firstpage}--\pageref{lastpage}}
\maketitle

\begin{abstract}
Unravelling galaxy formation theory requires understanding galaxies both at high and low redshifts. A possible way to connect both realms is by studying the oldest stars in the Milky Way (i.e., the proto-Galaxy). 
We use the \textsl{APOGEE-Gaia} surveys to perform a purely chemical dissection of Milky Way (MW) stellar populations, and identify samples of stars likely belonging to proto-Galactic fragments. 
The metallicity dependence of the distribution of old MW stars in the [Mg/Mn]-[Al/Fe] enables the distinction of at least two populations in terms of their star formation histories: a rapidly evolved population likely associated with the main progenitor system of the proto-MW; and populations characterised by less efficient, slower, star formation. In the Solar neighbourhood less efficient star forming populations are dominated by the \textsl{Gaia-Enceladus/Sausage} accretion debris. In the inner Galaxy, they are largely associated with the \textsl{Heracles} structure.
We model the density of chemically defined proto-Galaxy populations, finding that they are well represented by a Plummer model with a scale radius of $a\sim3.5$ kpc, and an oblate ellipsoid with flattening parameters [$p\sim0.8$; $q\sim0.6$]; this finding indicates that the MW plausibly hosts a low-mass, metal-poor, bulge component. We integrate this density for \textit{chemically unevolved} stars between $-2 < \mathrm{[Fe/H]} < -0.5$ to obtain a minimum stellar mass for the proto-Galaxy of $M_{*} (r<10~\mathrm{kpc}) = 9.1\pm0.2\times10^{8}$ M$_{\odot}$. Our results suggest the proto-Milky Way is at least comprised of two significant fragments: the main \textit{in situ} progenitor and the \textsl{Heracles} structure.
\end{abstract}


\section{Introduction} \label{sec:intro}

In order to fully understand galaxy formation theory, it is vital that we account for the data obtained at high and low redshift. The physical processes governing the genesis of galaxies during the earliest cosmic times set the scene for their subsequent evolution. From  the growth of galaxies and the formation of their discs and other Galactic components, to the birth and death of the first stars and their role in the production, enrichment, and dispersion of metals, many (if not all) processes in galaxy formation are shaped by the physics in the first few million years after the Big Bang. 

Thanks to the advancement in astronomical instrumentation over the past several decades, and the launch of the \textsl{James Webb Space Telescope} (JWST) \citep[]{jwst_2006, Jacobsen2022}, it is possible to study the high-redshift Universe in great detail. Several studies have looked into the distant past and have uncovered galaxies at high redshift (\citealp[$z\gtrsim8-10$;][]{Castellano2022,Robertson2023,Curtis2023, Carniani2024}) that have shed light into galaxy formation in the early Universe (\citealp[e.g.,][]{Inayoshi2022,Schaerer2022, Tacchella2022}).

In a similar vein, the high-redshift Universe can also be studied in our own Galaxy \footnote{Known as near-field cosmology.}. The oldest stars in the Milky Way are the relics of the physical processes in the early Cosmos \citep{Frebel_2015,Arentsen2020}. Thus, the properties of these stars retain the clues for understanding galaxy formation in this epoch. This is the main aim of Galactic archaeology: to decipher galaxy formation from the fossilised record encoded in Milky Way stars. However, in contrast to using imaging, photometry, and spectroscopy of whole galaxies like those supplied by telescopes like \textsl{JWST}, our ability to resolve stellar populations on a star-by-star basis in the Milky Way affords us the possibility to unravel the intricate processes of galaxy formation at high-redshift that is unrivalled by any other galaxy \citep[e.g.,][]{Arentsen2024}.

To that end, equipped with the revolutionary data from large-scale stellar surveys (primarily \textsl{Gaia}: \citealp{Gaia2022}, and \textsl{APOGEE}: \citealp{Majewski2017}), several studies have begun to unravel the earliest phases of formation in the Milky Way. For example, \citet{Horta2021} peered into the inner Galaxy using \textsl{APOGEE} DR16 and \textsl{Gaia} DR2 data and uncovered what seems to be the remnant of a major building block of the Galaxy, dubbed \textsl{Heracles}\footnote{One could reasonably speculate about the existence of a possible link between \textsl{Heracles} and globular cluster populations classified in terms of orbital parameters \citep{Massari2019} or age/metallicity \citep{Kruijssen2020}.  However, no such connections have been firmly established.}. \citet{Belokurov2022} analyzed metal-poor stars around the Solar neighbourhood using the synergy of \textsl{APOGEE} and \textsl{Gaia} data and identified a stellar population they called \textsl{Aurora}, that they claim is comprised of stars born \textit{in-situ} in the Milky Way proper. Using a much larger sample of stars with \textsl{Gaia} XP metallicities \citep{Andrae2023} in the inner Galaxy, \citet{Rix2022} clearly demonstrated that the Milky Way has a centrally concentrated overdensity of metal-poor old stars --- the Galaxy's ``\textit{poor old heart}'' \citep[see also][]{Chandra2023, Viswanathan2024}. All these lines of evidence suggest that there are several building blocks that are confined to the inner Galactic regions, are dominated by metal-poor stars on low net azimuthal velocities, with enhanced [$\alpha$/Fe] abundance ratios relative to solar. The amalgamation of these stellar populations is commonly dubbed the proto-Milky Way. Moreover, all these chemical-kinematic results qualitatively corroborate findings from theoretical studies examining cosmological simulations (\citealp[e.g.,][]{Horta2024_proto,Semenov2024}). 

While the insightful findings from these earlier studies have helped elucidate portions of the formation of the Galaxy at high-redshift, there is much that is yet still unknown. For example, a common problem that arises when examining metal-poor (stellar halo) populations in the Milky Way is defining the point in the metallicity distribution function (MDF) in which the Galactic disc ends and the Galaxy's stellar halo begins (\citealp[e.g.,][]{Belokurov2022,Conroy2022,Chandra2023,Zhang2024}); this alone can have direct implications on defining samples of stars comprising the most primordial populations in the Galaxy (i.e., the proto-Milky Way/proto-Galaxy), and in turn have bearing on the genesis of the Galactic disc \citep{Viswanathan2024}. Furthermore, currently there are no strong measurements of the structural/density profile of the proto-Milky Way (and its building block fragments), or how much stellar mass it amounts to (although see \citealt{Belokurov2023_nrich}). These quantities have strong ramifications on our understanding of the formation of classical bulges and the role of proto-Galaxy populations.

Further out in the stellar halo, between $5 < r < 30$ kpc, studies have shown that the density profile can be well modelled by a broken power law (\citealp[e.g.,][]{Deason2011,Whitten2019, Han2022, Amarante2024}) with a moderately shallow exponent ($\alpha\sim~2-4$) (\citealp[][]{Xue2015, Iorio2018, Mackereth2020}) and a break at $r\approx20-25$ kpc (\citealp[e.g.,][]{Deason2018, Deason2019, Han2022}). This density profile is conjectured to be associated with the distribution of the debris from the omniprescent \textsl{Gaia-Enceladus/Sausage} merger (\citealp[][]{Belokurov2018, Helmi2018, Deason2024}). We currently have a moderately clear picture of the structure and density profile of the outer regions of the stellar halo. However, this is not the case for the innermost regions, within $r\lesssim5$ kpc.

In this paper, we set out to synergise \textsl{APOGEE} (DR17) spectroscopic data with \textsl{Gaia} (DR3) astrometric data to take another step towards unravelling the earliest phases of the Milky Way, and in turn attempt to constrain galaxy formation at high-redshift. Specifically, we aim to further understand the density profile, the total mass, and the chemical-kinematic properties of stars belonging to the most primordial stellar populations comprising the founding building blocks of the Galaxy, and thus part of the proto-Milky Way (see Fig 1 from \citet{Horta2024_proto}, for example). To do so, we provide a detailed description of how to dissect stellar halo populations in chemical space and understand their kinematic and orbital properties (Section~\ref{sec:chemistry}) before modelling the (stellar) density of the proto-Milky Way (Section~\ref{sec:density}). We close by discussing our results in the context of previous work in Section~\ref{sec:discussion} before summarizing our conclusions in Section~\ref{sec:conclusions}.

\section{Data} \label{sec:data}

In this paper, we use a combination of spectroscopic data from the latest release of the \textsl{APOGEE} survey \citep[DR17][]{Majewski2017} and astrometric data from the third \textsl{Gaia} data release \citep[\textsl{Gaia} DR3][]{Gaia2022}. \textsl{APOGEE} data are based on observations collected by two high-resolution, multi-fibre spectrographs \citep{Wilson2019} attached to the 2.5m Sloan telescope at Apache Point Observatory \citep{Gunn2006} and the du Pont 2.5~m telescope at Las Campanas Observatory \citep{BowenVaughan1973}, respectively. Element abundances are derived using the \textsl{ASPCAP} pipeline \citep{Perez2015} based on the \textsl{FERRE} code \citep[][]{Prieto2006} and the line lists from \citet{Cunha2017} and \citet[][]{Smith2021}. The spectra themselves were
reduced by a customized pipeline \citep{Nidever2015}. For details on
target selection criteria, see \citet{Zasowski2013} for \textsl{APOGEE}, \citet{Zasowski2017} for \textsl{APOGEE}-2, \citet[][]{Beaton2021} for \textsl{APOGEE} north, and \citet[][]{Santana2021} for \textsl{APOGEE} south. Conversely, the \textsl{Gaia} mission/survey (\citealp[]{Gaia2016}) delivers detailed sky positions and proper motion measurements for $\sim2$ billion stars, limited only by their apparent magnitudes (\textsl{Gaia} $G\lesssim20.7$). Here we use only astrometric (positions and proper motions) measurements released in \textsl{Gaia} DR3 (\citealp[][]{Gaia2022}). 

We make use of the distances for the \textsl{APOGEE} DR17 catalogue generated by \citet{Leung2019b}, using the \texttt{astroNN} python package \citep[for a full description, see][]{Leung2019}. These distances are determined using a re-trained \texttt{astroNN} neural-network software, which predicts stellar luminosity from spectra using a training set comprised of stars with both APOGEE DR17 spectra and \textit{Gaia} EDR3 parallax measurements \citep{Gaia2020}.

The parent sample used in this work is comprised of stars that satisfy the following selection criteria:

\begin{enumerate}
    \item[\textbf{Red Giant Branch stars:}] \hfill \\ 
        \textsl{APOGEE}-determined atmospheric parameters, effective temperature and surface gravity, between $4000<$ $T_{\mathrm{eff}}$ $<5500$ K and $\log~g$ $< 3.5$,
    \item[\textbf{High signal-to-Noise spectra:}] \hfill \\ 
        \textsl{APOGEE} spectral S/N $> 50$,
    \item[\textbf{High-quality derived spectral parameters:}] \hfill \\ 
        \textsl{APOGEE} \texttt{STARFLAG} bits \textit{not} set to 0, 1, 3, 16, 17, 19, 21, 22, \textsl{APOGEE} \texttt{ASPCAPFLAG} bits \textit{not} set to 23, and \texttt{EXTRATARG} flag set to 0,
    \item[\textbf{Reliable distance measurements:}] \hfill \\
        $d/\sigma_{d} > 10$,
    \item[\textbf{No star clusters:}] \hfill \\ 
        Stars that are not within the \textsl{APOGEE} globular cluster value added catalogue \citep{Schiavon2023} or the catalogue from \citet{Horta2020},
    \item[\textbf{Reliable abundance measurements:}] \hfill \\ 
        Stars that have their \texttt{X\_FE\_FLAG} set to 0 for the following abundances: Mg, Al, Mn.
\end{enumerate}

All together, we use the 6-D phase space information\footnote{The positions, proper motions, and distances are taken/derived from \textit{Gaia} DR3 data, whilst the radial velocities are taken from \textsl{APOGEE} DR17.} and convert between astrometric parameters and Galactocentric cylindrical coordinates, assuming a solar velocity combining the proper motion from Sgr~A$^{*}$ \citep{Reid2020} with the determination of the local standard of rest of \citet{Schonrich2010}. This adjustment leads to a 3D velocity of the Sun equal to [U$_{\odot}$, V$_{\odot}$, W$_{\odot}$] = [--11.1, 248.0, 8.5] km s$^{-1}$. We assume the distance between the Sun and the Galactic Centre to be $R_{0}$ = 8.275~kpc \citep{Gravity2022}, and the vertical height of the Sun above the midplane $z_{0}$ = 0.02~kpc \citep{Bennett2019}. 

\begin{figure}
\centering
    \includegraphics[width=\columnwidth]{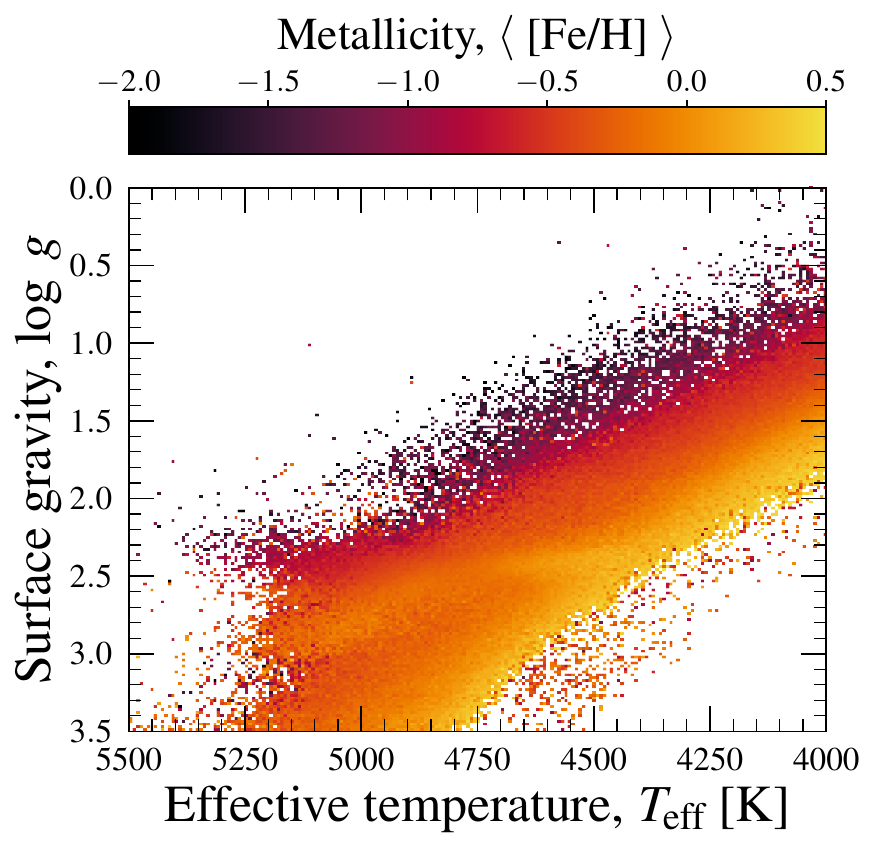}
    \caption{Kiel diagram for stars in our parent sample, binned into a 2D density distribution and color coded by the mean [Fe/H] in every pixel.}
    \label{fig:kiel} 
\end{figure}

\section{Chemically dissecting Milky Way stellar populations} \label{sec:chemistry}

\begin{figure*}
\centering
    \includegraphics[width=\textwidth]{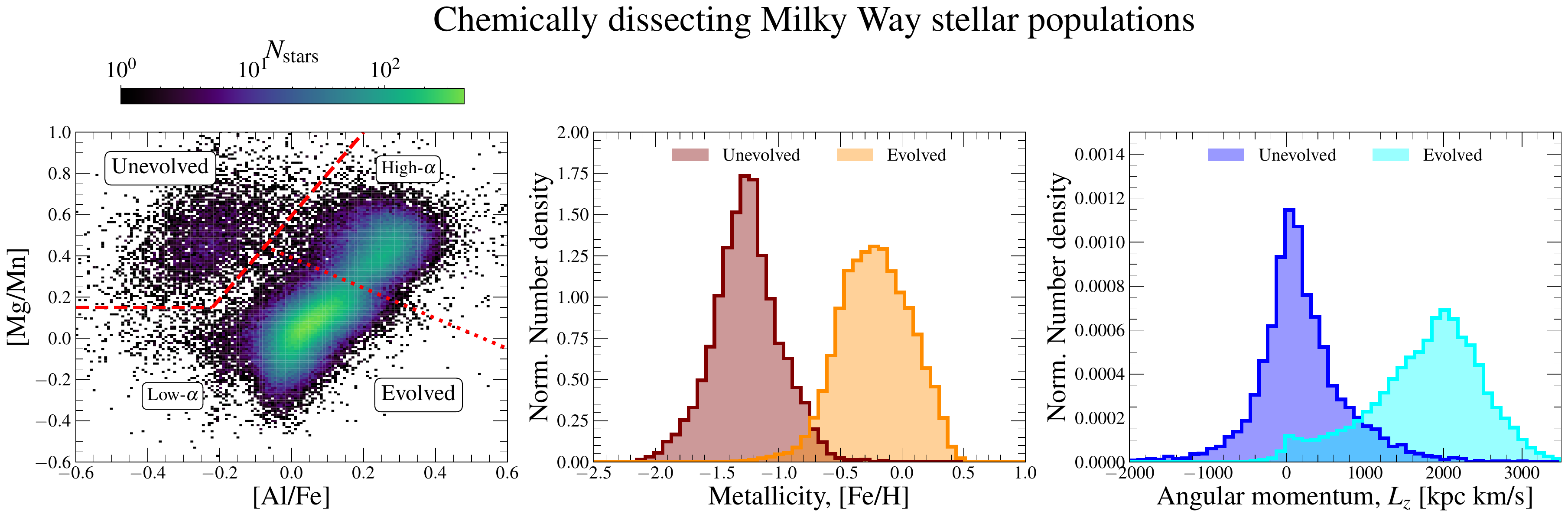}
    \caption{\textbf{Left:} 2D density distribution of stars in our parent sample in the [Mg/Mn]-[Al/Fe] plane. Here, we show with red dash(dotted) lines the division we use to select unevolved(high-/low-$\alpha$) populations. The slope of the line used to select unevolved stars in this diagram is given by [Mg/Mn] $\geq0.15~\cup$ [Al/Fe] $\leq-0.2~\wedge~$ [Al/Fe] $>-0.2~\cup~$ [Mg/Mn] $> 2\mathrm{[Al/Fe]} + 0.6$. \textbf{Center:} Normalised MDF of the chemically unevolved/evolved stars from. Here, chemically unevolved stars present a normal distribution centered around [Fe/H] $\sim~-1.3$, whereas evolved populations display a distribution centered around [Fe/H] $\sim~-0.2$. However, there is some clear overlap between these two samples, spanning from $-1.5 \lesssim $ [Fe/H] $\lesssim -0.5$. \textbf{Right:} Normalised distribution of the azimuthal angular momentum for unevolved/evolved stars. Similar to the MDF, both samples present vastly different distributions. Here, unevolved populations peak around $L_{z}\sim0$ kpc km s$^{-1}$, whereas evolved populations peak around $L_{z}\sim2\times10^{3}$ kpc km s$^{-1}$. However, both samples overlap substantially; unevolved populations reach up to $L_{z}\sim2\times10^{3}$ kpc km s$^{-1}$, and evolved populations reach down to $L_{z}\sim0$ kpc km s$^{-1}$. This illustrates that chemically unevolved stars can appear on highly rotational (disc-like) orbits and evolved stars can appear on non-circular (radial) orbits. Interestingly, we see a secondary peak in the chemically evolved sample around $L_{z}\sim0$ kpc km s$^{-1}$. This smaller peak likely corresponds to the metal-poor high-$\alpha$ sample \citep{Belokurov2022} and the kicked-up disc/\textsl{Splash} (\citealp{Bonaca2017,Belokurov2020}).}
    \label{fig:magnamal-intro} 
\end{figure*}

In the following, we perform an exploration of the data in chemistry space. We primarily discuss the [Mg/Mn] - [Al/Fe] chemical composition plane (Section~\ref{sec_magnamal}), initially introduced by \citet{Hawkins2015} and later modelled by \citet{Horta2021} (see also \citealt{Fernandes2023}).  This plane has been shown to be useful to distinguish stellar populations in the Milky Way (see also \citealp[][]{Das2020, Buder2022, Carrillo2022}). Our aim is to understand, using solely element abundance data, how different Galactic stellar populations behave in this plane; we also strive to decipher how the high-$\alpha$ disc and other metal-poor populations (i.e., accreted and $in$ $situ$ stellar halos) overlap in this key diagram. 

\begin{figure*}
\centering
    \includegraphics[width=\textwidth]{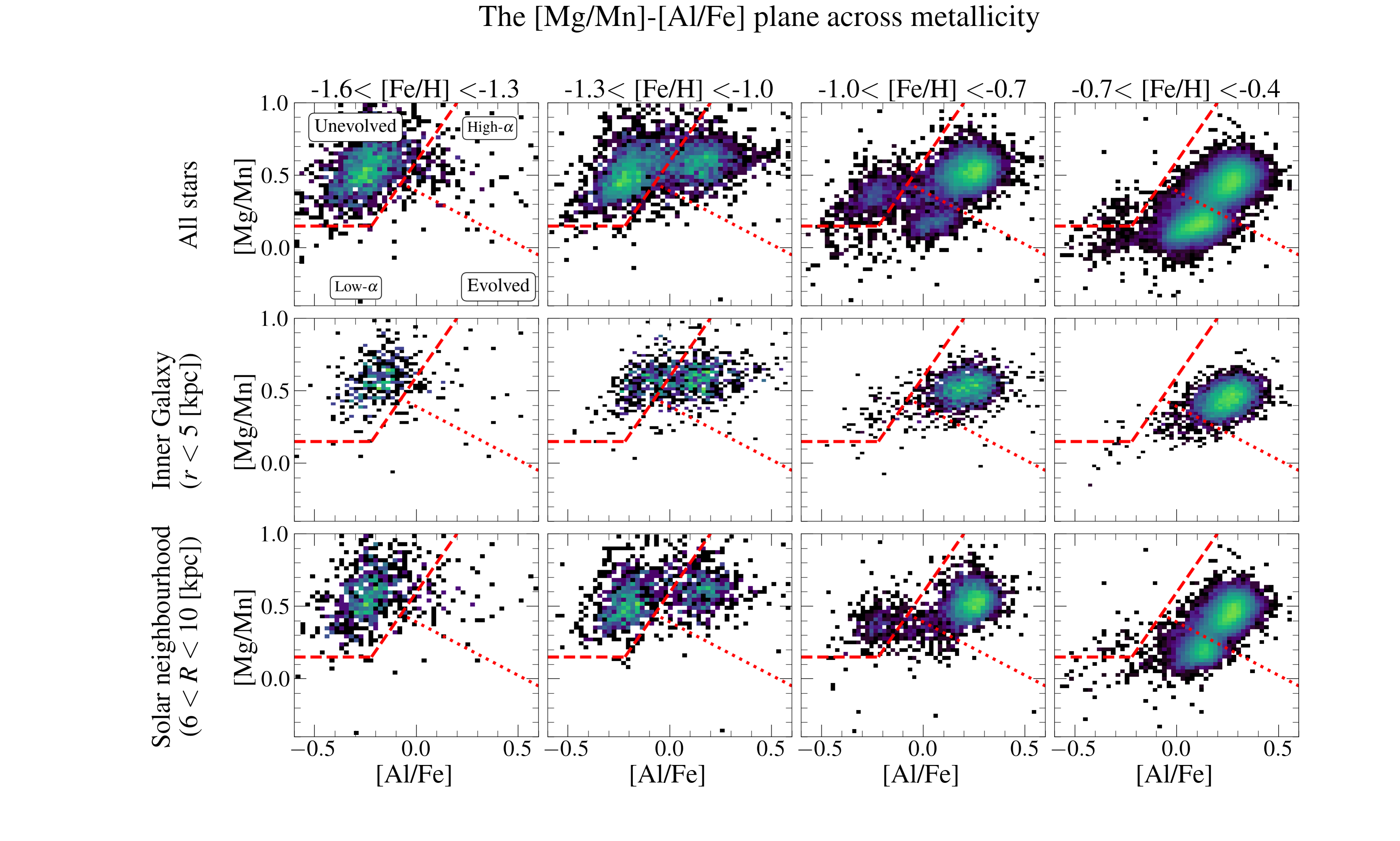}
    \caption{Sample of stars with [Fe/H] $< -0.4$ from our parent sample in the [Mg/Mn]-[Al/Fe] plane, but now binned in [Fe/H]. Each bin has a width of 0.3 dex. At high [Fe/H] (rightmost panel), the high-/low-$\alpha$ sequences are dominant, and there is barely any unevolved populations (although possibly some residue of \textsl{GES} debris at low [Mg/Mn]). Between $-1.6 < $ [Fe/H] $< -0.7$, there is a trend observed where evolved(unevolved) populations decrease(increase) in numbers with decreasing [Fe/H]. The low-$\alpha$ sequence disapears below [Fe/H] $\approx-1$, and the unevolved populations appear at [Fe/H] $\approx-0.7$. Interestingly, the high-$\alpha$ sequence appears to host stars all the way down to [Fe/H] $\sim-1.6$. This result indicates that a selection in the [Mg/Mn]-[Al/Fe] plane is {\it not} equivalent to a hard cut in [Fe/H], and illustrates that high-$\alpha$ populations still inhabit the high-$\alpha$ region of the diagram at low [Fe/H]. See Section~\ref{sec_magnamal} for further details.}
    \label{fig:magnamal-fehs} 
\end{figure*}

\subsection{The [Al/Fe]-[Mg/Mn] chemical plane}
\label{sec_magnamal}

The left panel of Fig~\ref{fig:magnamal-intro} shows a 2D density distribution of the parent sample used in this work in the [Mg/Mn]-[Al/Fe] plane. Marked in this diagram as a dashed (dotted) red line are two boundaries we impose to define unevolved-evolved (high-$\alpha$-to-low-$\alpha$) populations, following the definitions adopted by \citet{Horta2021}. We stress that these boundaries are far from arbitrary, but are used to demarcate three already clearly defined loci occupied by the data: the unevolved population, the high-$\alpha$ population, and the low-$\alpha$ population. 

The middle and right panels of Fig~\ref{fig:magnamal-intro} show the MDF and angular momentum distribution, $L_{z}$, of the unevolved and evolved (high-$\alpha$ $+$ low-$\alpha$) populations. As can be seen from Fig~\ref{fig:magnamal-intro}, the selection of unevolved/evolved stellar populations in the [Mg/Mn]-[Al/Fe] plane does $not$ lead to a hard cut in metallicity. Instead, it yields two clear distributions. The one peaked at ${\rm [Fe/H] \approx -1.3}$ corresponds to the unevolved component, whereas the other peak at ${\rm [Fe/H] \approx -0.2}$ corresponds to the evolved population (including both high- and low-$\alpha$ discs).  Interestingly, the two distributions overlap between ${\rm -1.5 \simless [Fe/H] \simless -0.5.}$  

Similarly, when looking at the distribution of $L_{z}$, we find that again there are two clear peaks, one at $L_{z} \sim 2,000$ kpc km s$^{-1}$ (evolved) and the other at $L_{z} \sim 0$ kpc km s$^{-1}$ (unevolved), with some considerable overlap between the two peak values.  
In fact, unevolved populations can reach high $L_{z}$ orbits, meaning that some stars can follow near-circular orbits resembling disc populations. However, the majority of the stars in the unevolved population are on low $L_{z}$ (radial) orbits. This is because the sample of unevolved stars in \textsl{APOGEE-Gaia} is dominated by the debris of the \textsl{Gaia-Enceladus/Sausage} (GES) accretion event \citep[][]{Belokurov2018,Helmi2018, Haywood2018, Mackereth2019b}, that is dominated by stars on highly radial orbits. Moreover, there are also some stars on higher retrograde orbits ($L_{z}\lesssim-1,000$ kpc km s$^{-1}$), which could make up a different part of the debris from the GES \citep[][]{Horta2023} or debris from distinct mergers \citep[e.g.,][]{barba2019,Myeong2019}. Interestingly, we find that (chemically) evolved populations can present low angular momentum orbits. These reach down to $L\sim0$ kpc km s$^{-1}$, in line with the more classical definition of \textit{in situ} stellar halo populations \citep[e.g.,][]{Eggen1962,Beers2012}. As we show in the following, these stars are likely part of the metal-poor high-$\alpha$ sequence (or \textsl{Aurora} population: \citealp[][]{Belokurov2022}) and the heated disc/\textsl{Splash} population \citep{Bonaca2017,Belokurov2020}.

In order to visualise how stellar populations in the [Mg/Mn]-[Al/Fe] plane change as a function of metallicity, in Fig~\ref{fig:magnamal-fehs} we show our parent sample (top) in different bins of [Fe/H], each 0.3 dex wide, spanning from $-1.6$ to $-0.4$. We also show separate plots focusing on those stars in the inner Galaxy (middle, $r<5$ kpc) and those in the Solar neighbourhood (bottom, $6<R<10$ kpc). 
We first focus on the top panels of Fig~\ref{fig:magnamal-fehs}.  As expected from the middle and right panels of Fig~\ref{fig:magnamal-intro}, one can see that: 1) stars located within the high-$\alpha$ region of the diagram appear at metallicities below [Fe/H] $<-1.3$, all the way down to [Fe/H] $\approx-1.6$. These metal-poor high-$\alpha$ stars are part of the metal-poor extension of the high-$\alpha$ sequence, and also likely part of the \textsl{Aurora} population \citep{Belokurov2022}; 
2) unevolved populations appear in metallicity bins above [Fe/H] $=-1$, and their position in the unevolved region of the diagram varies with [Fe/H]. They are situated in a position of lower [Mg/Mn] at [Fe/H] $>-1$, and increase in [Mg/Mn] with decreasing [Fe/H], following approximately the chemical evolution path characteristic of low mass satellites of the Milky Way\footnote{This is to be expected, as the contribution by SN type Ia decreases towards lower metallicity, leading to a higher [Mg/Mn].}\citep[see][for an more detailed study]{Fernandes2023}; 
3) there is a spur of stars in the $-0.7<$ [Fe/H] $<-0.4$ bin at [Mg/Mn] $\sim0$ and [Al/Fe] $<-0.1$. These stars are only seen in the Solar neighbourhood, and are likely part of the most metal-rich GES stars that fall outside our boundary of unevolved populations. However, we cannot rule out the possibility of these stars being low-$\alpha$ disc stars with peculiar [Al/Fe] abundances \citep{Feuillet2022}.
Interestingly, the number of high-$\alpha$ stars drops with decreasing metallicity, reaching down to [Fe/H] $\sim -1.6$ in the Solar neighbourhood. In the inner Galaxy the evolved populations also reach down to low metallicity, albeit slightly higher than [Fe/H] $\sim-1.6$.

The metallicity dependence of the high-$\alpha$ star distribution in this plane merits consideration.  The direction of chemical evolution in the [Mg/Mn]-[Al/Fe] plane, and in particular the horizontal path connecting the ``unevolved’’ and the high-$\alpha$ loci, is highly dependent on the star formation history/efficiency (see Fig 6 from \citealt{Fernandes2023} for example). 
The existence of a population of stars with relatively high [Al/Fe] at metallicities as low as [Fe/H] $\sim -1.6$ suggests that the \textit{in situ} population formed during a highly efficient star formation period, causing a quick horizontal evolution between the ``unevolved'' and the high-$\alpha$ regions. 
In that scenario, the \textit{in situ} stars inhabiting the ``unevolved’’ region of the [Mg/Mn]-[Al/Fe] plane should be considerably and predominantly metal-poor (below [Fe/H] $\lesssim-1.6$, see \citealp[][]{Arentsen2024}). It is noteworthy, however, that within both the inner Galaxy and the Solar neighbourhood, star counts surge towards [Fe/H] $<-1$ in the unevolved region. 

Given the MDF of the stellar populations in the unevolved region of the [Mg/Mn]-[Al/Fe] plane, the apparent dependence of [Mg/Mn] on metallicity in those populations (see also Fig. 3 from \citealt{Horta2021} and Fig. 5 of \citealt{Fernandes2023}), and in view of the above reasoning, the surplus of stars seen in the ``unevolved'' region with metallicities between  $-1.6<$ [Fe/H] $<-0.7$ must be comprised predominantly by the debris of systems \textit{not} formed in the main progenitor of the Milky Way.
In the Solar neighbourhood, this population of chemically unevolved stars is dominated by the \textsl{Gaia-Enceladus/Sausage} debris (\citealp[][]{Helmi2018,Belokurov2018,Das2020}), whereas in the inner Galaxy, these stars largely comprise the debris of the \textsl{Heracles} structure \citep{Horta2021}, and possibly other less massive structures. 
We hypothesize that in both systems, because of a lower star formation rate, Al-enrichment takes place at a much slower pace, making possible the build up of a sizeable population substantially enriched by SNIa (i.e., with higher [Fe/H]), that is characterised by relatively low [Al/Fe] \citep{Fernandes2023}. We note that our conclusions are not changed by implementation of the chemical definitions of accreted vs {\it in situ} stars adopted by \cite{Belokurov2022}. However, we also note that at metallicities below $-1.6$ dex, it is not possible to disentangle chemically unevolved populations with the current data.

In summary, analysis of the distribution of populations of different metallicity on the [Mg/Mn]-[Al/Fe] plane unveils the presence of ``evolved'' high-$\alpha$ stars as metal-poor as $\rm [Fe/H] \approx -1.6$ (Fig~\ref{fig:magnamal-fehs}).  This finding indicates that the {\it in situ} population underwent a fast evolution in [Al/Fe], roughly along the lines of the chemical evolution models presented by \cite{Horta2021}.  
Consequently, any {\it in situ} populations present in the ``unevolved'' locus of that plane are predominantly and considerably metal-poor, typically with ${\rm [Fe/H]\simless-1.6}$. Since the MDF of stars in our sample in that region peaks at much higher metallicity (middle panel of Fig.~\ref{fig:magnamal-intro}), it follows that the ``unevolved'' stellar population contains a sizeable population comprised of stars not formed in the Milky Way main progenitor.  In the Solar Neighbourhood, those typically belong to the \textsl{Gaia-Enceladus/Sausage} system. In the inner Galaxy, they belong mostly to \textsl{Heracles}.  This is an important conclusion, as it suggests that it is possible to distinguish populations formed within and outside of the main Milky Way progenitor for stars with [Fe/H]$>-1.6$, both in the inner Galaxy and elsewhere, on the basis of precision elemental abundance measurements and the principles of chemical evolution theory.

\subsection{Vetting chemically selected populations in kinematic space}
\begin{figure*}
\centering
    \includegraphics[width=\textwidth]{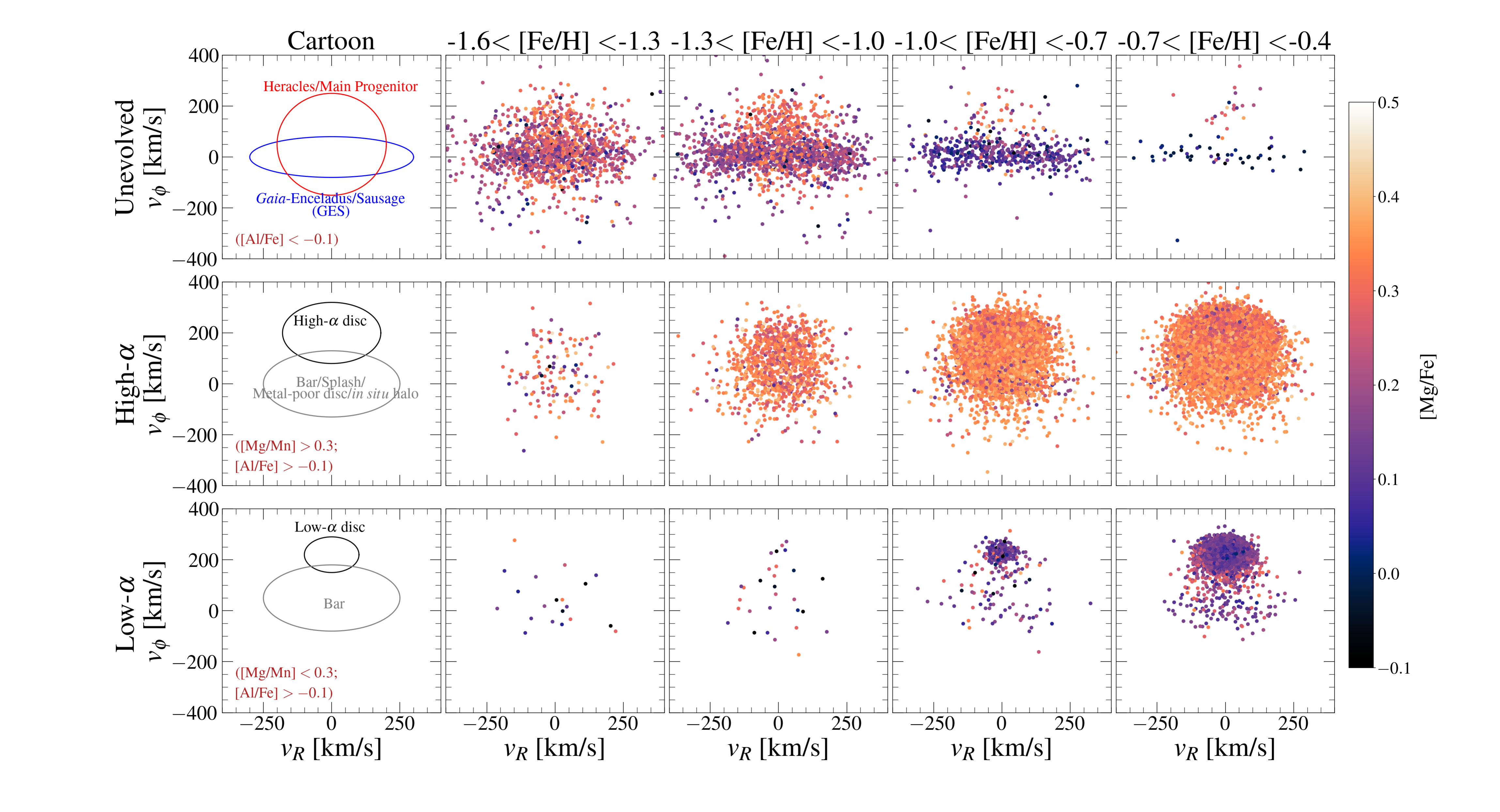}
    \caption{Azimuthal velocity of stars, $v_{\phi}$, as a function of their radial velocity, $v_{R}$. Each row includes the different locii of stars from the [Mg/Mn]-[Al/Fe] plane (top is unevolved, middle is high-$\alpha$, bottom is low-$\alpha$). Each row is then divided into four different panels binned by [Fe/H], going from [Fe/H] $=-1.6$ (left) to [Fe/H] $=-0.4$ (right), each spanning 0.3 dex. In the leftmost column, we also show a cartoon illustration of where different stellar populations sit in these diagrams, to guide the eye. The main takeaway result is that chemically unevolved populations are dominated by two distinct distributions: the radial GES debris (purple), and the isotropic \textsl{Heracles}/main progenitor populations (orange). See Section~\ref{sec:chemistry} for further details in the main text.}
    \label{fig:magnamal-vphi-vR} 
\end{figure*}

With the selection of unevolved/evolved stellar populations from Fig~\ref{fig:magnamal-intro}, we go on to study how these populations appear in velocity-[Fe/H] space. Fig~\ref{fig:magnamal-vphi-vR} shows three rows, where the top/middle/bottom row shows the unevolved/high-$\alpha$/low-$\alpha$ populations from Fig~\ref{fig:magnamal-intro}, respectively. Each row is then divided into five panels, splitting the populations by [Fe/H], as done in Fig~\ref{fig:magnamal-fehs}. In each of these panels, we show the azimuthal velocities, $v_{\phi}$, for each star as a function of their (cylindrical) radial velocities, $v_{R}$; each star is colour-coded by their [Mg/Fe] abundance ratio. Lastly, to guide the eye, in the leftmost column we show a cartoon of where different stellar populations are expected to sit in this diagram for each sample. 

In the top row (unevolved), one can see how there are two clear populations in this diagram: an elongated population that has lower-[Mg/Fe] and reaches higher $v_{R}$ (the \textsl{GES} debris), and a round population with higher [Mg/Fe], centred at $[v_{R}, v_{\phi}]$ $\sim [0, 50]$ km s$^{-1}$ (the \textsl{Heracles}/proto-Milky Way populations). The \textsl{Heracles}/proto-Milky Way stars appear in all diagrams, but are predominantly seen between $-1.6<$ [Fe/H] $<-1$. Conversely, the GES debris appear more dominant at [Fe/H] $>-1.3$, and reach [Fe/H]$\sim-0.5$. 

In the second(third) row, the high-$\alpha$(low-$\alpha$) evolved populations are shown. Here, stars at [Fe/H] $>-1$ and $v_{\phi}>100$ belong to the high-/low-$\alpha$ discs, respectively. However, there is a population of stars in both the high-$\alpha$ and low-$\alpha$ samples that present no net azimuthal rotation ($v_{\phi}\sim0$ km s$^{-1}$). For the high-$\alpha$ population ---that are characterized by [Al/Fe] $\gtrsim-0.1$---, these stars are comprised of an amalgamation of distinct stellar populations: the innermost disc/Galactic bar, the heated disc (or \textsl{Splash}, \citealp{Bonaca2017,Belokurov2020}), and the \textsl{Aurora} population \citep[][]{Belokurov2022}. The differences between these populations are subtle, and are mostly due to their different typical values of [Fe/H] and spatial distributions in the Galaxy\footnote{The more metal-rich bar and inner disc are within $R\lesssim4$ kpc, whilst the more metal-poor \textsl{Splash} stars are located around the Solar neighbourhood (see \citet{Belokurov2020} for details). Metal-poor \textsl{Aurora} stars, although identified in \citet{Belokurov2022} around the Solar neighbourhood, are also conjectured to be present in the inner Galaxy.}. Similarly, in the low-$\alpha$ sample, the low-$\alpha$ disc is dominant at [Fe/H] $>-1$. However, there are a number of low-$\alpha$ stars with [Al/Fe] $\gtrsim-0.1$ at all  [Fe/H] that show low azimuthal velocities ($v_{\phi}\sim0$ km s$^{-1}$). For these stars that have [Fe/H] $>-1$, we conjecture that they are likely part of the bar or recently discovered \textsl{knot} component in the innermost Galaxy \citep[][see also \citealp{Rix2024}]{Horta2024_knot}. For the more metal-poor populations we conjecture they are likely GES stars that have infiltrated our low-$\alpha$ sample. However, we cannot rule out the possibility of these stars comprising a heated (low-$\alpha$) disc (\textsl{Splash}) population.

\begin{figure}
\centering
    \includegraphics[width=\columnwidth]{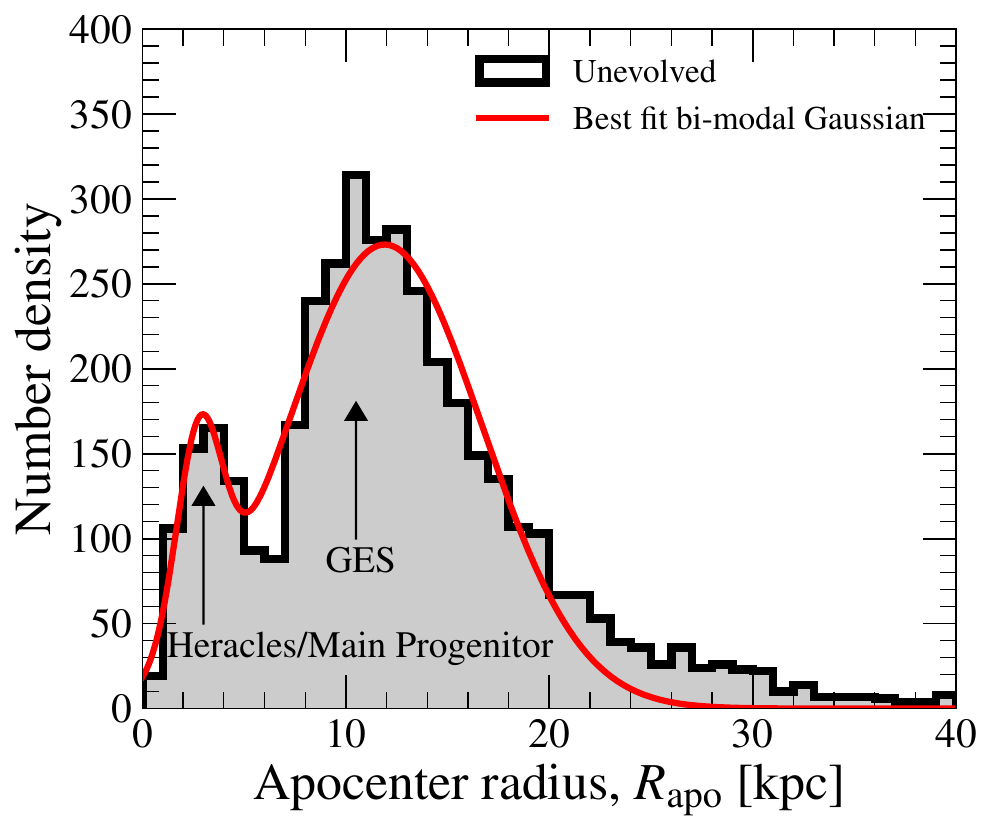}
    \caption{Apocenter radii distribution for chemically unevolved stars selected in Fig~\ref{fig:magnamal-intro}. In red we also show the best fit bi-modal Gaussian to the data. The chemical selection yields a distribution that is bimodal, with a peak at $R_{\mathrm{apo}}\sim5$ kpc (\textsl{Heracles}/main progenitor) and another at $R_{\mathrm{apo}}\sim12-15$ kpc (GES debris). This could be due to the selection function of the \textsl{APOGEE} survey. However, in Fig~\ref{fig:magnamal-vphi-vR} we see that the chemically unevolved stars appear to show a bimodal distribution in $v_{\phi}$-$v_{R}$, whereby the GES debris appear very radial and present lower [Mg/Fe], at fixed [Fe/H], than the \textsl{Heracles}/main progenitor populations, that appear more isotropic and present higher [Mg/Fe]. This result suggests that the two peaks in $R_{\mathrm{apo}}$ are distinct (see Fig~\ref{fig:Tinsley}).}
    \label{fig:rapo} 
\end{figure}

\begin{figure*}
\centering
    \includegraphics[width=\textwidth]{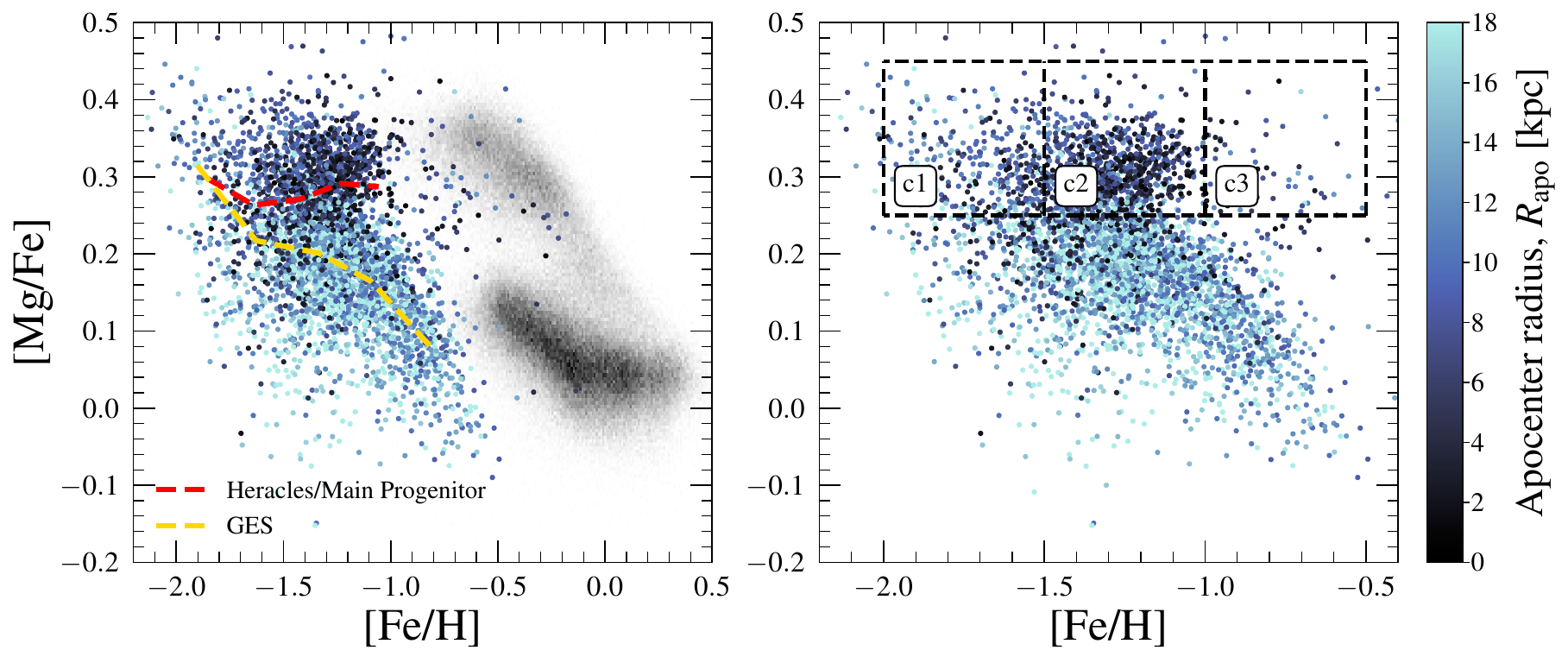}
    \caption{\textbf{Left}: [Mg/Fe]-[Fe/H] diagram of the chemically unevolved stars, colored by their apocenter radius. We also show the parent sample as a 2D histogram in the background. Plotted also are the running medians for stars with $R_{\mathrm{apo}}<5$ kpc (red) and $R_{\mathrm{apo}}>8$ (yellow), to guide the eye. There are two clear sequences in this diagram, one that is traced by the \textsl{GES} debris (yellow), and one that is traced by the \textsl{Heracles}/main progenitor populations (red). \textbf{Right:} Same as left, but now with the chemical cell grid we impose to sub-select chemically unevolved populations; cells 1-3 are those that contain predominantly \textsl{Heracles}/proto-Milky Way populations.}
    \label{fig:Tinsley} 
\end{figure*}

\subsection{Selecting sub-samples of unevolved stars in chemical cells}

Interestingly, if one isolates the chemically unevolved stars from Fig~\ref{fig:magnamal-intro} and plots their apocenter radius, $R_{\mathrm{apo}}$, distribution (Fig~\ref{fig:rapo}), two distinct peaks appear\footnote{Apocenter radii are computed integrating orbits using the latest version of the \texttt{gala} software package (\citealp{Price2017,gala2022}), assuming the \texttt{MilkyWayPotential2022} potential.}. The dominant peak is centered around $R_{\mathrm{apo}}\sim12-15$ kpc, and corresponds to the \textsl{GES} debris. Conversely, the other peak is centered around $R_{\mathrm{apo}}\sim5$ kpc, thus confined to the inner Galactic regions, and comprises the \textsl{Heracles}/main progenitor populations. We note that while these two peaks are distinct, these two distributions overlap; the \textsl{Heracles}/main progenitor populations likely extend to $R_{\mathrm{apo}}\sim12$ kpc and the GES debris reach $R_{\mathrm{apo}}\sim5$ kpc. However, the relative strengths of the two peaks should not be interpreted quantitatively as these are a result of the \textsl{APOGEE} selection function.

\begin{figure}
\centering
    \includegraphics[width=\columnwidth]{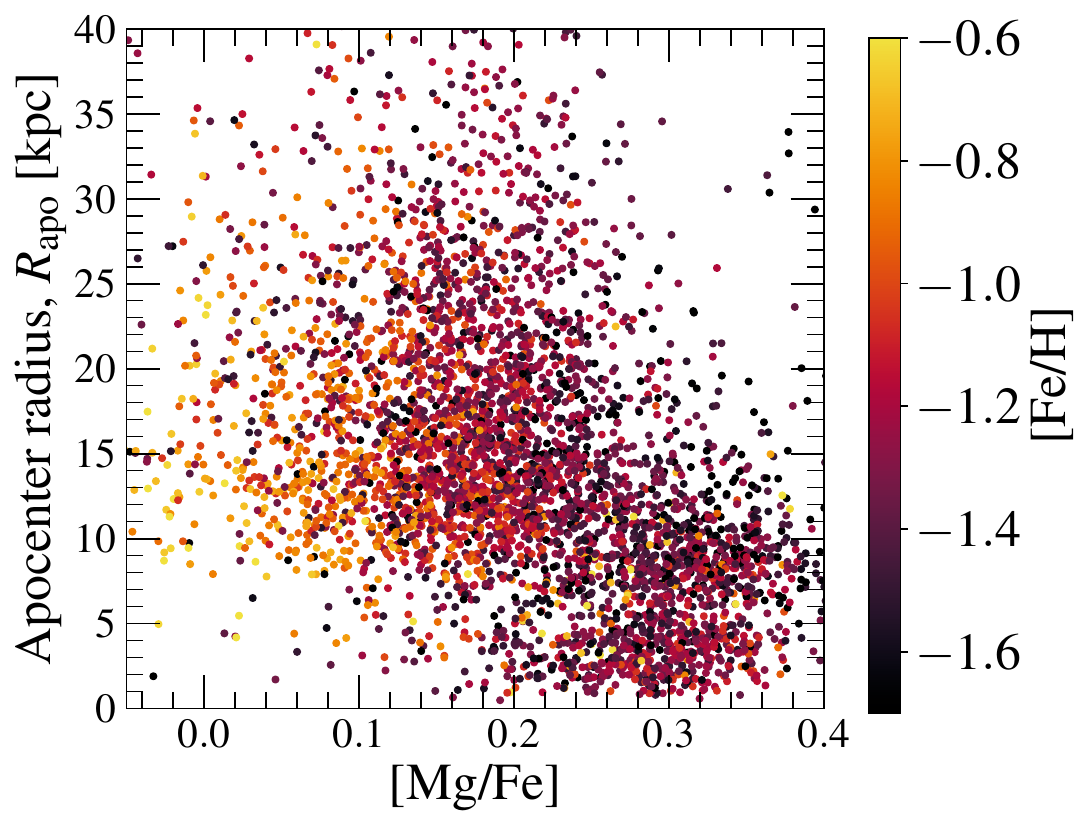}
    \caption{Apocenter radius vs [Mg/Fe] abundances for chemically unevolved stars. There are two distinct populations in this diagram that overlap in metallicity: a more tightly concentrated, $R_{\mathrm{apo}}\lesssim12$ kpc, population with higher $\alpha$-to-iron abundances, [Mg/Fe]$\sim0.35$ (i.e., \textsl{Heracles}/main progenitor); a less confined distribution, $8\lesssim~R_{\mathrm{apo}}\lesssim30$ kpc, with overall lower average $\alpha$-to-iron abundances, [Mg/Fe]$\sim0.18$ (i.e., the \textsl{Gaia-Enceladus/Sausage}).}
    \label{fig:rapo-mgfe} 
\end{figure}

The fact that we see two clear distributions in $R_{\mathrm{apo}}$ space when looking at chemically unevolved populations supports the result obtained in Fig~\ref{fig:magnamal-vphi-vR} when looking in kinematic space; the chemically unevolved sample hosts two superimposed populations that appear to have different kinematic/orbital properties. As we saw in Fig~\ref{fig:magnamal-vphi-vR}, there is also a difference in these populations based on their [Mg/Fe], where the \textsl{Heracles}/main progenitor population appears more enchanced than the GES debris at fixed [Fe/H] \citep{Horta2021,Horta2023}. This can be seen even more clearly in Fig~\ref{fig:Tinsley}, where we show the unevolved stars in the [Mg/Fe]-[Fe/H] diagram colored by their $R_{\mathrm{apo}}$. For completeness, we also show the parent sample as a 2D density distribution in the background. The two populations we have seen in both [Mg/Fe] (Fig~\ref{fig:magnamal-vphi-vR}) and $R_{\mathrm{apo}}$ (Fig~\ref{fig:rapo}) trace different trajectories in this plane, providing further evidence that these two populations are distinct. If the bimodal distribution seen in Fig~\ref{fig:rapo} (or in orbital energy space) was caused by the footprint of the \textsl{APOGEE} selection function, as suggested by \citet{Lane2022} and \citet{Myeong2022}, we would expect these two populations to trace the same (chemical) trajectory in Fig~\ref{fig:Tinsley}. 

Moreover, the same result is also illustrated in Fig~\ref{fig:rapo-mgfe}, where we show the [Mg/Fe] vs $R_{\mathrm{apo}}$ distribution of chemically unevolved stars coloured by their [Fe/H] value. There are two distinct structures in this diagram that overlap in metallicity: a more tightly concentrated, $R_{\mathrm{apo}}\lesssim12$ kpc, population with higher $\alpha$-to-iron abundance ratios, [Mg/Fe]$\sim0.35$ (i.e., \textsl{Heracles}/main progenitor); and a less confined distribution, $8\lesssim~R_{\mathrm{apo}}\lesssim30$ kpc, with overall lower average $\alpha$-to-iron abundances, [Mg/Fe]$\sim0.18$ (i.e., the \textsl{Gaia-Enceladus/Sausage}). We note that the stripe of missing stars at $R_{\mathrm{apo}}\approx8$ kpc seen in the \textsl{Heracles}/main progenitor (higher [Mg/Fe]) population is likely a projection of the \textsl{APOGEE} selection function.

Given these subtle but key differences, it is possible to select distinct sub-samples of chemically unevolved populations using solely chemistry. We do this by laying down a grid of chemical cells (or mono-abundance populations, MAPs: \citealp{Bovy2016}), as shown in the right panel of Fig~\ref{fig:Tinsley}. We perform this chemical selection of sub-samples of unevolved populations in order to then model their density distribution, accounting for the \textsl{APOGEE} selection function, independently. Our aim is to isolate, without the need for any spatial, kinematic, or orbital selection, stars associated with the \textsl{Heracles}/main progenitor population, in order to then determine the density profile and shape of this sample, and obtain an estimate of its stellar mass. In addition, we also set out to model the density of stars in all these cells simultaneously.

We note that due to the strong overlap between the \textsl{GES} debris and the \textsl{Heracles}/proto-Milky Way populations in all chemical-kinematic/orbit diagrams, there is bound to be some cross-contamination. 
Assuming that the density of \textsl{GES} debris in this plane is constant across $R_{\mathrm{apo}}$ and $v_{R}$ space, we assess this contamination by estimating the ratio of stars with $R_{\mathrm{apo}}>12$ kpc and $|v_{R}|>150$ km~s$^{-1}$ in each cell with the rest of the stars in that cell (i.e., $R_{\mathrm{apo}}<12$ kpc and $|v_{R}|<150$ km~s$^{-1}$), as stars with these orbital/kinematic properties are likely to be part of the GES debris (Fig~\ref{fig:magnamal-vphi-vR} and Fig~\ref{fig:rapo-mgfe}); we find a contamination fraction from GES to our samples in Cells (1, 2, 3) of ($16\%$, $11\%$, $3\%$), respectively.

In the following section, we describe the density modelling procedure, and set out to model unevolved stellar populations in the chemical cells.

\section{Density modelling} \label{sec:density}

We describe the method used to perform a modelling of the underlying number density of red giants in the Milky Way from \textsl{APOGEE} observations in units of stars per kpc$^{-3}$, $\nu_{*}(X,Y,Z|\theta)$. The computation of this quantity requires taking into account: 1) the (pencil-beamed nature) survey selection function of \textsl{APOGEE}; 2) inhomogeneous dust extinction along the lines of sight; 3) the target selection criteria imposed by different $H$-band magnitude limits; 4) the use of red giants as tracers that are not standard candles. In detail, we aim to model the spatial distributions of chemically selected stellar populations in the Milky Way to determine the density profile governing these samples, as well as the amount of (stellar) mass comprising each population.

\subsection{Density fitting procedure}
To perform the density modelling procedure, we use the publicly available \texttt{apogee} code \citep{apogeecode_2016}, that has been shown to accurately model the spatial density profiles of Milky Way stellar populations (\citealp[][]{Bovy2016, Mackereth2017,Mackereth2020, Horta2021_nrich,Lane2022, Imig2023}). The full details of the method can be found in \citet{Bovy2016}. However, we present here a summarised version for completeness. 

Under the assumption that star counts are well modelled by an inhomogeneous Poisson point process that takes into account the \textsl{APOGEE} selection function, we fit the observed sample of red giant branch (RGB) stars from \textsl{APOGEE} using a given density functional form (see Section~\ref{density-profiles} for details). Here, we consider stars to be distributed in space defined by $\mathcal{O} = [l, b, D, H, [J-K_{s}], \mathrm{[Fe/H]}]$, with an expected rate, $\lambda(\mathcal{O}|\theta)$, where $\theta$ is the vector of parameters in the density model of the rate function that are to be determined. In detail, the rate function can be expressed fully as

\begin{multline}
\label{eq:rate}
    \lambda(\mathcal{O}|\theta) = \nu_{*}(X, Y, Z|\theta) \times |J(X, Y, Z; l, b, D)|\\
    \times \rho(H, [J-K_{s}], \mathrm{[Fe/H]}|X, Y, Z)
    \times S(l, b, H, [J-K_{s}]),
\end{multline}
where $\nu_{*}(X, Y, Z|\theta)$ is the stellar number density in rectangular coordinates (units of stars per kpc$^{-3}$), $|J(X, Y, Z; l, b, D)|$ is the Jacobian of the transformation from rectangular ($X, Y, Z$) to Galactic ($l, b, D$) coordinates, $\rho(H, [J-K_{s}], \mathrm{[Fe/H]}|X, Y, Z)$ represents the density of stars in magnitude, colour, and metallicity space given a spatial position ($X, Y, Z$) ---in units of stars per arbitrary volume in magnitude, colour, and metallicity space---, and $S(l, b, H, [J-K_{s}])$ is the survey selection function \citep[see][for details]{Bovy2016, Mackereth2020}, that denotes the fraction of stars successfully observed in the survey's colour and magnitude range, including dust extinction effects. During this process, our method is able to correct for effects induced by interstellar extinction using combined 3D dust maps of the Milky Way derived by \citet{Marshall2006} for the inner disc plane and those derived for the majority of the \textsl{APOGEE} footprint by \citet{Green2019} (adopting conversions $A_{H}/A_{K_{s}} = 1.48$ and $A_{H}/E(B-V) = 0.46$, \citealp[][]{Schlafly2011, Yuan2013}). 

Once the rate function is estimated, under the assumption that the rate, $\lambda(\mathcal{O}|\theta)$, only depends on $\theta$ through $\nu_{*}(X, Y, Z|\theta)$, the log-likelihood of the Poisson point process can be be expressed as 

\begin{equation}
    \mathrm{ln}~\mathcal{L}(\theta|\mathcal{O}_{i}) = \sum_{i}^{N}\bigg[\mathrm{ln}~\nu_{*}(X_{i}, Y_{i}, Z_{i}|\theta) - \mathrm{ln}~\int{\mathrm{d}\mathcal{O}\lambda(\mathcal{O}|\theta)\bigg]}.
\end{equation}

The integral in the log-likelihood function describes the effective observable volume of the survey, which for \textsl{APOGEE} is a sum of integrals over each field in the survey; this can be expressed as

\begin{multline}
    \int{d\mathcal{O}\lambda(\mathcal{O}|\theta}) = \sum_{\mathrm{fields}}\Omega_{f}\int{\mathrm{d}DD^{2} \nu_{*}[X, Y, Z](D, \mathrm{field}|\theta)}\\
    \times \mathcal{S}(\mathrm{field}, D), 
\end{multline}

where $\nu_{*}[X, Y, Z](D, \mathrm{field})$ is the density, as before, but evaluated along each line of sight in an \textsl{APOGEE} field and $\mathcal{S}(\mathrm{field}, D)$ is the spatial effective selection function (see \citet{Mackereth2020} for further details).

Once a density model functional form is chosen, we optimize the likelihood function for a given stellar population sample using a downhill-simplex algorithm. We then use the initialized set of parameters $\theta$ to initiate a Markov Chain Monte Carlo (MCMC) sampling of the posterior probability distribution function (PDF) of the parameters in the density law using the affine-invariant ensemble sampler implemented in the python package \texttt{emcee} \citep{Foreman2013}. We then adopt the [$16^{th}, 50^{th}, 84^{th}$] percentiles as our median and standard deviation of one-dimensional projections of the MCMC chains as our best-fitting parameter values and uncertainties. For the uncertainty on values derived from these parameters, such as the stellar mass, we use the [$16^{th}, 84^{th}$] parameter percentile values from these posterior samples and estimate the mass integrating the density with these parameter values.

\subsection{Density profiles}\label{density-profiles}

Cells 1, 2, and 3 (Fig~\ref{fig:Tinsley}) are comprised by chemically unevolved stars that are primarily confined to the inner Galactic regions ($R_{\mathrm{apo}}\lesssim10-12$ kpc), and present higher [Mg/Fe] (at fixed [Fe/H]) than the \textsl{GES} debris. As shown in \citet{Horta2021} and \citet{Horta2023}, these stars belong to the \textsl{Heracles} population, but likely also other proto-Galaxy fragments. As there has been little amount of work modelling the density of metal-poor stars in the innermost regions of the Galaxy, we test several functional density forms. Broadly, these density models fall into the following categories: an exponential disc with $h_{R}=2.2$ kpc and $h_{z}=0.8$ kpc \citep{Mackereth2017, Horta2024}, a cored power-law profile, an Einasto profile, and a few other profiles that are commonly used to describe the density of elliptical galaxies and galaxy bulges, the Plummer \citep{Plummer1911}, Hernquist \citep{Hernquist1990}, and a S\'ersic profile \citep{Sersic1963}. For all of these models (except the exponential disc), we also test a version of each model allowing for triaxiality (i.e., $r_{e} = X + \frac{Y}{p} + \frac{Z}{q}$), where $p$ and $q$ are the flattening parameters. Lastly, we also test fitting these stars with the halo model from \citet{Mackereth2020}, who modelled the density of high eccentricity stars ($e>0.7$) in the Milky Way stellar halo with \textsl{APOGEE} data using a complex form of a triaxial single power-law profile (see Appendix~\ref{appen_fits} for details). In total, we fit the stars in cells 1, 2, and 3 with 12 different density profiles. For each model we compute the maximum negative log-likelihood, the Bayesian Information Criterion (BIC), and the Akaike information criterion (AIC), to assess the best fitting model for each chemical cell from Fig~\ref{fig:Tinsley}. The description of each density profile can be found in Appendix~\ref{appen_fits}, whilst the resulting maximum negative log-likelihood, BIC, and AIC values are shown in Table~\ref{tab:bic}. Lastly, we adopt flat priors for all parameters in all models; here, any scale radius parameter has bounds between [0, 50], the parameters related to the slopes of a power law have bounds between [0, 30], and the parameters relating to triaxiality and disc contamination have bounds between [0, 1].

After testing several different models and assessing the likelihood, BIC, and AIC values (see Appendix~\ref{appen_fits} for details), we find that the best fitting model for cells 1, 2, and 3 is the triaxial Plummer model (Fig~\ref{fig:best-fits-all}). This is also the case when modelling all stars in all three cells at once. Thus, we use this model, and the parameters from the optimized likelihood function, to perform the MCMC sampling of the PDF of the parameters in the Plummer model. In detail, the triaxial Plummer model takes the following form:

\begin{equation}
    \nu_{*}(r_{e}) \propto \frac{3}{4\pi a^{3}} \Bigg(1+\frac{r_{e}^{2}}{a^{2}}\Bigg)^{-5/2},
\end{equation}
where $r_{e} = X + \frac{Y}{p} + \frac{Z}{q}$ and $a$ is the Plummer radius. The resulting best profile fits are described in Table~\ref{tab:fit-values}.

\subsection{Mass estimates}

In order to estimate the mass for the chemically unevolved stars in different chemical cells, we use the best-fitting model and its associated uncertainty for a distribution of stars. With these numbers in hand, the mass can be computed by employing the normalisation of the rate function (Eq~\ref{eq:rate}).

We perform this mass-estimation procedure for stars in chemical cells 1--3, demarked in Fig~\ref{fig:Tinsley}, by calculating the number of stars seen by \textsl{APOGEE} for a given density model normalised to unity at the Solar position, \textit{N}(\textit{$\nu_{*,0}$} = 1). That number is obtained by integrating the rate function over the observable volume of the survey. This integral is given by:
\begin{multline}\label{eq8}
        N(\nu_{*,0} = 1) = \int_{\mathrm{fields}} d \textup{field}\, dD\, \lambda(\mathrm{field},D) = \\
        \int d \textup{field}\, d\mu\,\frac{D^{3}(\mu)\mathrm{log}(10)}{5}\nu_{*}([R,\phi,z](\mathrm{field},\mu|\tau))\\ \times \mathcal{S}(\mathrm{field},\mu),
\end{multline}
where the density and effective selection function (namely, $\mathcal{S}$(field,$\mu$)) are calculated along \textsl{APOGEE} sightlines on a grid linearly spaced in distance modulus $\mu$. Since the
true number of observed stars is given by:
\begin{equation}
{N}_{\mathrm{obs}} \,=\, {A\, N}(\nu_{*,0}=1),
\end{equation}
comparison of the expected number count for a normalised density model with the
true observed number of stars in the sample provides the proper
amplitude, $A$, which is then equivalent to the true number density of
RGB stars at the Sun, $\nu_{*,0}$.

The number counts in RGB stars can be converted into the mass of the entire underlying population. To do so, we use the \textsl{PARSEC} isochrones (\citealp{Bressan2012,Marigo2017}), weighted with a \citet{Kroupa2001} IMF. The average mass of RGB stars $\langle$$M_{\mathrm{RGB}}$$\rangle$ observed in \textsl{APOGEE} is then calculated by applying the same cuts in (\textit{J} -- $K_{s}$)$_{0}$ and $\log{g}$ to the isochrones. The fraction of the stellar mass in giants, $\omega$, is given by the ratio between the IMF weighted sum of isochrone points within the RGB cuts and those of the remaining population. Using this fraction, the conversion factor between giant number counts and total stellar mass can be determined using:
\begin{equation}\label{eq9}
    \chi(\mathrm{[Fe/H]}) = \frac{\langle M_{\mathrm{RGB}} \rangle (\mathrm{[Fe/H]})}{\omega(\mathrm{[Fe/H]})}.
\end{equation}

As explained in \citet{Mackereth2020}, the factor for each field and each selection in [Fe/H] is computed by adjusting the limit in ($J$ -- $K_{s})_{0}$ to reflect the minimum ($J$ -- $K_{s})_{0}$ of the bluest bin adopted in that field, and only considering isochrones that fall within the metallicity limits given by each chemical cell. The edges in colour binning for each field are accounted for by the integration over $\rho$[($J$ -- $K_{s})_{0}$,$H$] for the effective selection function. This factor can be as large as $~600$ $M_{\odot}$ star$^{-1}$ for the lowest [Fe/H] bins, in fields where the \textsl{APOGEE} ($J$ -- $K_{s})_{0}$ limit was 0.5. For higher metallicity bins, and in fields where \textsl{APOGEE} assumed a bluer ($J$ -- $K_{s})_{0}$ $> 0.3$ cut, this number approaches $~200$ $M_{\odot}$ star$^{-1}$. \citet{Mackereth2020} used Hubble Space Telescope photometry to show that the factors determined using their method are reliable against systematic uncertainty arising from the choice of stellar evolution models. Combining these factors with the number density normalization, we attain the appropriate mass normalization, $\rho_{0} = \chi(\mathrm{[Fe/H]}) \times \nu_{*, 0}(\mathrm{[Fe/H]},\mathrm{[Mg/Fe]})$, for each cell. Finally, upon determination of the normalisation of a sample, we integrate the normalised density models described by 400 samples from the posterior distributions of their parameters to attain the total mass within a population. 

\subsection{Modelling the density of proto-Milky Way populations}

\begin{figure*}
\centering
    \includegraphics[width=\textwidth]{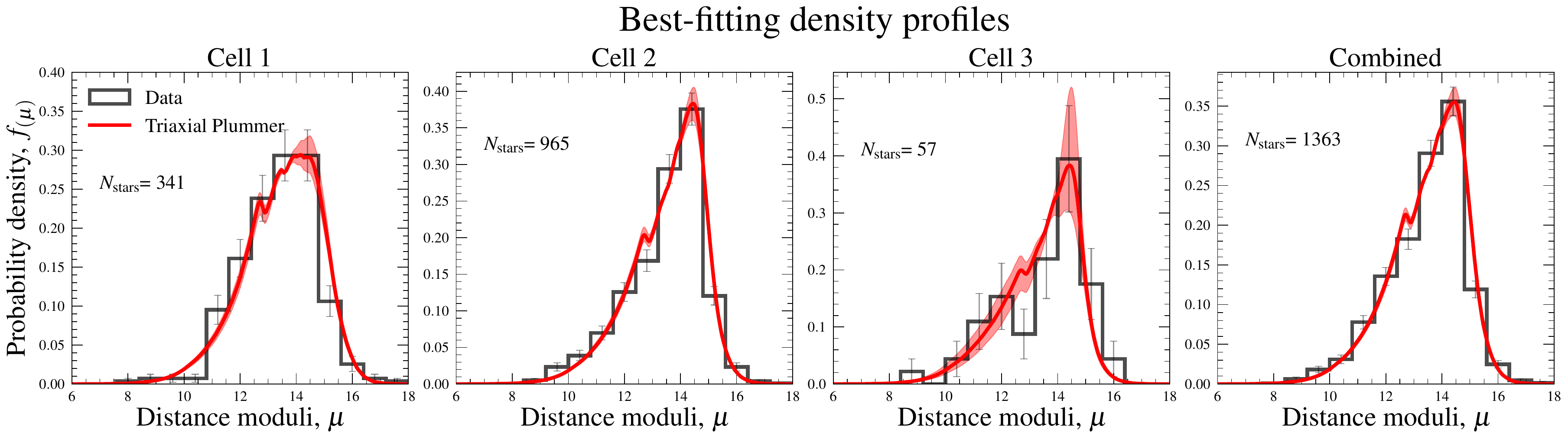}
    \caption{Best-fit density profiles (Triaxial Plummer models) for the chemical cells containing stars belonging to \textsl{Heracles}/main progenitor populations compared to the data, as well as the best fit model to all data in all three cells. The error bars on the histograms illustrate the Poisson error in every bin.}
    \label{fig:best-fits} 
\end{figure*}

Fig~\ref{fig:best-fits} shows the best fitting density profile compared to the data for cells 1--3 (as well as combined), that contain stars belonging to proto-Galactic fragments. We find that the stars in these cells are best fit by a triaxial Plummer profile (see Fig~\ref{fig:best-fits}, Table~\ref{tab:fit-values}, and Table~\ref{tab:bic}). All these cells present flattening parameters $p$ and $q$ that are non-zero, meaning that the shape of the ellipsoid is triaxial. The median values of $p$ are [0.82, 0.77, 0.95] for cells 1, 2, and 3, respectively, and the combined measurement is $p=0.8\pm0.03$. These results imply that there is little flattening in the $y$ direction. Moreover, we find that the median values of the vertical flattening parameter $q$ are [0.78, 0.63, 0.41] for cells 1, 2, and 3, respectively, and the combined result is $q=0.66\pm0.02$. Thus, there is clearly more flattening in the $z$ direction (w.r.t. the present day Milky Way disc). This indicates that the overall distribution is closer to an oblate ellipsoid. We find that the median values of the Plummer radius are [5.1, 3.1, 2.5] (kpc) for cells 1, 2, and 3, respectively, and $a=3.48\pm0.10$ when modelled all together. The Plummer radius is a scale parameter that sets the size of the core. Thus, it is interesting that we find that the data are well described by a model that is tightly concentrated in the inner Galaxy. 
Interestingly, we find that within the uncertainties, the scale radius and $q$ flattening parameters do not agree between cells, despite the values being very close. We reason that this discrepancy is likely due to the higher contamination from GES debris in Cell 1 (see Section~\ref{sec:chemistry} for details), which would lead to a more extended profile (and thus larger scale radius) that is less flattened in the $z$ direction.

We integrate the density within 10 kpc to estimate the amount of stellar mass in each cell, and find that cell 1 amounts to $M_{*} = 1.7\pm0.1 \times 10^{8}$ M$_{\odot}$, cell 2 amounts to $M_{*} = 7.2\pm0.2 \times 10^{8}$ M$_{\odot}$, and cell 3 amounts to $M_{*} = 0.5\pm0.2 \times 10^{8}$ M$_{\odot}$. Summed up all together, we estimate a stellar mass for the high-$\alpha$ sequence of chemically unevolved stars (i.e., \textsl{Heracles}/main progenitor) of $M_{*} = 9.5\pm0.2 \times 10^{8}$ M$_{\odot}$. If modelled all together, the resulting mass estimate we obtain is $M_{*} = 9.1\pm0.2 \times 10^{8}$ M$_{\odot}$. Within the uncertainties, the mass we obtain from summing up all three cells and the mass obtained when modelling all three cells are in agreement.

In a final remark, we note that although this is not the main focus of the paper, we also tested running several different models for fitting the density of stars belonging to the \textsl{Gaia-Enceladus/Sausage} debris (not shown). We found that the best fitting models ranged depending on whether we included an eccentricity cut ($e>0.7$: \citealp[][]{Mackereth2020}) or not (see also \citealp[][]{Lane2023} for a more in depth discussion on the kinematic selection function in \textsl{APOGEE}). However, these stars could be well modelled by a triaxial single power-law as in previous studies \citep{Iorio2018,Mackereth2020, Lane2023}, or several other profiles (e.g., Einasto, Hernquist, cored power-law, Triaxial Plummer).

\begin{table*}
	\centering
	\caption{From left to right: chemical cell modelled containing stars belonging to the proto-Galaxy, best fit density profile for stars in that cell, posterior median and [16,84]$^{\mathrm{th}}$ percentiles of the best-fit parameters for the used model, integrated stellar mass. Measurements of the Plummer radius, $a$, are in units of kiloparsec. Estimates of the stellar mass are integrated within $r<10$ kpc.}
	\begin{tabular}{lcccccr} 
		\hline
		 Chemicall cell $\#$ & Density profile & Best-fit parameters & Stellar mass (M$_{\odot}$)\\
   \hline
    1 & Triaxial Plummer & $a=5.07^{+0.30}_{-0.30}$; $p=0.82^{+0.07}_{-0.07}$; $q=0.78^{+0.05}_{-0.05}$ & $1.7\pm0.1~\times~10^{8}$\\
    2 & Triaxial Plummer & $a=3.11^{+0.10}_{-0.10}$; $p=0.77^{+0.03}_{-0.04}$; $q=0.63^{+0.03}_{-0.03}$& $7.2\pm0.2~\times~10^{8}$\\
    3 & Triaxial Plummer & $a=2.50^{+0.30}_{-0.40}$; $p=0.95^{+0.03}_{-0.07}$; $q=0.41^{+0.05}_{-0.06}$& $0.5\pm0.2~\times~10^{8}$\\
    Combined & Triaxial Plummer & $a=3.48^{+0.10}_{-0.10}$; $p=0.80^{+0.03}_{-0.03}$; $q=0.66^{+0.02}_{-0.02}$& $9.1\pm0.2~\times~10^{8}$\\
\hline
	\end{tabular}
 \label{tab:fit-values}
\end{table*}

\section{Discussion} \label{sec:discussion}

\subsection{The interface between the Milky Way disc and stellar halo}

This work aims to assess the density profile of primoridal Milky Way populations (i.e., the proto-Galaxy), and to determine an estimate of its mass. To do so, it is imperative that high-purity populations of primordial Milky Way stars are selected, reducing contamination from disc or any other populations. Moreover, such selection must be free of any phase-space cuts in order to not bias the density modelling fits. 

In Section~\ref{sec:chemistry}, we performed an exploration of the \textsl{APOGEE-Gaia} data with the aim of determining a high-purity sample of chemically unevolved stars, like those expected for proto-Galaxy populations. We have resorted to using the [Mg/Mn]-[Al/Fe] plane (Fig~\ref{fig:magnamal-intro}), that has been shown to be effective at dissecting halo/unevolved populations from disc/evolved ones (\citealp[e.g.,][]{Hawkins2015, Das2020, Horta2021}) in the \textsl{APOGEE} data (although see \citet{Buder2022} for an alternative with Na). The power of this chemical diagram comes from the odd-Z element Al in the x-axis. Al is postulated to be an element whose yields are metallicity dependent \citep{Nomoto2013}, so [Al/Fe] is typically found to be depleted in dwarf galaxies relative to Galactic disc populations at fixed [Fe/H] (\citealp[e.g.,][]{Hasselquist2021,Horta2023}). Hence, the oldest stars in the Galaxy (proto-Galaxy) and stars formed in dwarf galaxies accreted onto the Milky Way, should present a lower [Al/Fe] ratio than those of Milky Way disc populations, assuming all stars form chemically from the same initial conditions. 

The time in which different systems evolved from this depleted [Al/Fe] valley into the higher [Al/Fe] region of Fig~\ref{fig:magnamal-intro} depends both on the star formation efficiency and (baryonic) mass of the system (see \citealt{Fernandes2023}). However, all stellar populations should start their chemical evolution in the unevolved region of the diagram (see Fig~2 from \citealt{Horta2021}). Thus, the key question for those interested in identifying the oldest stars formed in the Milky Way is: ``\textit{where does the disc end and the stellar halo begin?}''.

\citet{Belokurov2022} used [Al/Fe] and [Fe/H] cuts to divide metal-poor stars in the \textsl{APOGEE-Gaia} data into high-/low-[Al/Fe] (namely, at [Al/Fe]$\sim-0.1$), that they associate with accreted/$in$ $situ$ populations, respectively. When compared to the results in this work, the [Al/Fe] cut in that work would approximately equate to our selection of high-$\alpha$ and unevolved populations (Fig~\ref{fig:magnamal-intro}). They adopted this division to study the evolution of the azimuthal velocity, $v_{\phi}$, of [Al/Fe] $>-0.1$ metal-poor stars to measure the spin-up of the Milky Way disc; from this analysis they concluded that the metal-poor high-$\alpha$ sample spun-up into the Galactic disc approximately between $-1.3 < $[Fe/H] $<-0.9$. Our results from Fig~\ref{fig:magnamal-vphi-vR} (middle row) are fully consistent with that proposition, allowing for a wider metallicity transition region, as they show that the metal-poor high-$\alpha$ sample transitions from a non-rotating population to a highly rotational one at low metallicity (see also \citealp[][]{Viswanathan2024}).  This is not surprising, since both analysis are based essentially on the same data.

As a consequence of the results from \citet{Belokurov2022}, one would not expect any stars stars between $-1.6<$ [Fe/H]$<-1$ formed \textit{in situ} to inhabit the chemically unevolved region of Fig~\ref{fig:magnamal-intro}. This is because, under the assumption that all stellar populations begin in the chemically unevolved region of the diagram (see Fig 2 from \citealt{Horta2021}), any star formed \textit{in situ} would have to be more [Fe/H]-depleted than the most metal-poor high-$\alpha$ stars, whose lowest metallicity is [Fe/H] $\sim-1.6$ (see Fig~\ref{fig:magnamal-fehs} and \citealp[][]{Arentsen2024})\footnote{Recall that all of the metal-poor \textit{in situ} stars from \citep{Belokurov2022} have [Al/Fe] $\gtrsim-0.1$ by definition, and thus sit in the high-$\alpha$ locus.}.
Fig~\ref{fig:magnamal-fehs} reveals a different picture, however. When examining the all stars sample (top row), we see that there is a clear overdensity in the chemically unevolved region of the diagram that extends to [Fe/H] $\approx-0.7$. This is also the case if we split the parent sample into inner Galaxy stars (middle row), and stars around the Solar neighbourhood (bottom row). For the Solar neighbourhood sample, these stars inhabiting the chemically unevolved region of the diagram belong to the \textsl{Gaia-Enceladus/Sausage} debris (\citealp[see also][]{Das2020}). Conversely, for the inner Galaxy sample, these chemically unevolved stars belong to the \textsl{Heracles} system \citep{Horta2021}, and possibly some additional proto-Galaxy populations. Thus, the surplus of chemically unevolved stars in the inner Galaxy with metallicities between $-1.6 \lesssim$ [Fe/H] $\lesssim-1$ is evidence that the Milky Way hosts a sizeable stellar population in its inner region, whose star formation history must have been different than that of the main progenitor component of the proto-Galaxy. \citet{Horta2021} ascribes the majority of this population to a massive building block called \textsl{Heracles}. 

We note here that we have been careful to label the evolved region of the diagram as high-/low-$\alpha$ regions, and \textit{not} high-/low-$\alpha$ \textit{disc} regions. This is because if one examines the kinematics of these different (chemically selected) populations (Fig~\ref{fig:magnamal-vphi-vR}), we see that high-$\alpha$ evolved stars can still present kinematics with low $v_{\phi}$, and thus not highly rotating. There could be a variety of reasons why these stars present low $v_{\phi}$ values: their orbits have been affected by a merger (\citealp[][]{Bonaca2017, Belokurov2020}), they are on orbits caught on resonances (\citealp[][]{Dillamore2024}), they are part of the metal-poor high-$\alpha$ sequence typically assumed to be linked to the stellar halo (\citealp[][]{Carollo2010, Haywood2018, Dillamore2019}), they are part of the primordial \textit{in situ} population \citep[i.e., \textsl{Aurora}:][]{Belokurov2022}. However, these results indicate that the metal-poor high-$\alpha$ sample, that should be more chemically evolved than populations comprising the proto-Galaxy, appears to extend to metallicities well below [Fe/H]$<-1.3$. This result would lead to conclude that the oldest and most primordial stars in the Galaxy must present [Fe/H] values that fall below the [Fe/H] of the high-$\alpha$ sequence ([Fe/H] $\lesssim-1.6$, \citealp[][]{Arentsen2024}), assuming all these stars formed in the same system. 

In summary, our results suggest that selecting stars in the [Mg/Mn]-[Al/Fe] plane provides a more accurate path to detemine a high-purity sample of halo/unevolved stars than in [Al/Fe]-[Fe/H]. A selection in [Al/Fe]-[Fe/H] yields a sample of stars that inhabit the high-$\alpha$ region of the diagram, and thus likely are not comprised primarily by the stars formed in the proto-Galaxy. This is because [Fe/H] can overlap for stars with different [Al/Fe]. Moreover, the fact we see stars in the unevolved region of the diagram ---where we expect proto-Galaxy populations to inhabit--- at metallicities up to [Fe/H]$\sim-1$ that present different kinematics and [$\alpha$/Fe] abundance ratios (top row of Fig~\ref{fig:magnamal-vphi-vR}) to the GES debris supports the hypothesis of additional building blocks existing in the heart of the Galaxy \citep[\textsl{Heracles}:][]{Horta2021}. This would lead to conclude that the proto-Milky Way is comprised of many building blocks, and not solely one main progenitor. This has been predicted by cosmological simulations \citep{Horta2024_proto} and has been shown to be the case in recent \textsl{JWST} observations \citep{Mowla2024}.

\subsection{The density and mass of the proto-Galaxy}

Given our purely chemical selection of unevolved stellar (halo) populations, we have set out to model those stars that are likely a mix of \textsl{Heracles} and the proto-Milky Way. These stars sit in the chemically unevolved region of the [Mg/Mn]-[Al/Fe] diagram, present higher [Mg/Fe] ---at fixed [Fe/H]--- than the GES debris, and are confined to the inner Galaxy ($R_{\mathrm{apo}}\lesssim5$ kpc, Fig~\ref{fig:rapo}, but extend to $R_{\mathrm{apo}}\sim12$ kpc, Fig~\ref{fig:rapo-mgfe}). We have selected these stars by gridding chemically unevolved stellar populations in the [Mg/Fe]-[Fe/H] plane, and selecting stars in different chemical cells (Fig~\ref{fig:Tinsley}). We have then tested several different density functional forms, and have assessed the best fitting profile (Fig~\ref{fig:best-fits}). We have found that stars in all three cells are well modelled by a Plummer model with a core radius of approximately $\sim3-5$ kpc, that is predominantly oblate in shape, with axis ratios parameters on the order of $p\sim0.8-0.95$ and $q\sim0.4-0.8$. All together, we have estimated a stellar mass for all three cells of $M_{*} = 9.5\pm0.2 \times 10^{8}$ M$_{\odot}$ within $r<10$ kpc. When modelling the data from all three cells together, we get a combined measurement of $a = 3.48$ kpc, $p=0.80$, $q=0.66$, and $M_{*} = 9.1\pm0.2 \times 10^{8}$ M$_{\odot}$.

The fact that we find the best fitting profile to be a triaxial Plummer model may not be that surprising. Plummer models have been shown to be useful at describing the stellar density profile of spherical/ellipsoidal systems (e.g., galaxy bulges and globular clusters, \citealp[][]{Plummer1911}). The Plummer model also falls in the family of density models able to describe the profile of dark matter halos in dwarf galaxies \citep[e.g.,][]{Wilkinson2002}. Recent studies modelling cosmological simulations seem to suggest that the oldest and most centrally concentrated stars in the Galaxy follow the distribution of the dark matter, regardless of their accretion history and environment \citep{lucey2024}. Stars from cells 1--3 are well modelled by a Plummer model, but can also be modelled to decent accuracy using models commonly used to describe dark matter halos: e.g., Einasto (see Appendix~\ref{appen_fits}). Furthermore, the value obtained for the vertical flattening parameter $q<1$ implies that the stellar populations comprising the proto-Galaxy manifest an oblate shape. This is also what is expected for massive proto-Galaxy populations given recent results from cosmological simulations \citep{Horta2024_proto}. The value of $p\sim0.8-0.95$ implies that there is little evidence for flattening in the $Y$ direction, alike the further out ($r\gtrsim10$ kpc) stellar halo (\citealp[$p\sim0.8$: e.g.,][]{Deason2019,Mackereth2020, Han2022, Lane2023}). 

Most of the stellar mass of inner Galaxy populations ($r<5$ kpc) is postulated to be comprised by the boxy-peanut bulge (\citealp[e.g.,][and references therein]{Ness2012,Ness2016,Portail2017}\footnote{Although see also the recent discovery of a knot in the Milky Way's center (\citealp{Horta2024_knot, Rix2024}).}), that are more confined to the Galactic midplane, following the Milky Way bar. In modelling the density profile of chemically unevolved stars likely belonging to populations comprising the proto-Galaxy, we have revamped the question of the existence of a bulge component in the Milky Way. Our results suggest that this is potentially the case, with a scale radius parameter on the order of $a\sim3.5$ kpc. However, this component only amounts to a small fraction of the total mass in this region ($\sim5\%$).

Lastly, in modelling the density of stars belonging to proto-Galactic fragments, we estimated a mass for this system. Integrating the density for stars in cells 1--3 yields a total mass of $M_{*}=9.1\pm0.2 \times 10^{8}$ M$_{\odot}$ \footnote{We note that this mass should be treated as a lower bound measurement, as \textsl{APOGEE} does not probe well stars below [Fe/H] $\lesssim-1.6$, where most of the stars belonging to the main progenitor of the Milky Way are expected to be.} within $r<10$ kpc. This estimate is within the regime of the expected mass for massive building blocks and/or proto-Galaxy populations that predict anywhere between $ 2 \times 10^{8} \lesssim M_{*} \lesssim 5 \times 10^{9}$ M$_{\odot}$ \citep{Horta2024_proto}. 
Moreover, if added to the current estimate of the Milky Way's stellar halo mass ($\sim10^{9}$ M$_{\odot}$; \citealp[][]{Deason2019}), this would double the current estimate of the stellar halo,  $M_{*}\approx2\times10^{9}$ M$_{\odot}$.

Interestingly, if one breaks up each cell, stars in chemical cell 2 are primarily comprised by the \textsl{Heracles} debris \citep{Horta2021}, which occupies a primary locus within $-1.5 <$ [Fe/H] $< -1$ and $0.25 <$ [Mg/Fe] $<0.45$. From our estimates, we find that cell 2 amounts to a mass of $M_{*}=7.2\pm0.2 \times 10^{8}$ M$_{\odot}$; this is a little higher than what was initially predicted in \citep{Horta2021}, who used the mean [Mg/Fe]-[Fe/H] abundance ratios to estimate a mass of $M_{*}\sim5 \times 10^{8}$ M$_{\odot}$ by comparing to cosmological simulations. We argue that this difference can be attributed to the data and simulations not being a perfect 1--1 comparison. Furthermore, it could also be the case that Cell 2 hosts a small fraction of proto-Milky Way populations that don't belong to \textsl{Heracles}, that would lead to a higher mass estimate. Regardless, \textsl{Heracles} is likely a building block system with a mass that is comparable to that of the main progenitor system of the proto-Milky Way. If this scenario is correct, \textsl{Heracles} must have caused a big impact on the formation of the Milky Way. 

However, it is also the case that we are not sampling the main branch progenitor component of the proto-Galaxy with the \textsl{APOGEE-Gaia} data, as: $i$) we are not probing well the low-metallicity end of the MDF, at [Fe/H] $<-2$, where the data is scarce. Studies examining [Fe/H]$\lesssim-2$ stars in the inner Galaxy \citep[e.g.,][]{Arentsen2020, Arentsen2023, Arentsen2024} could aid in this effort; $ii$) because of our definition of chemically unevolved populations, we are not including the most metal-rich stars ($-1.6\lesssim$[Fe/H]$\lesssim-1.3$: \citealp{Belokurov2022}) belonging to the main progenitor in our density modelling fit. Recent estimates from examining the distribution of destroyed globular cluster populations suggest this is on the order of $\sim5\times 10^{8}$ M$_{\odot}$ \citep{Belokurov2023_nrich}.

\subsection{Connection to high-redshift Galaxy formation}

Recent studies using \textsl{JWST} have suggested the discovery of the earliest stages of mass assembly of a Milky Way-mass progenitor in action (the \textsl{Firefly Sparkle}: \citealp[e.g.,][]{Mowla2024}). Such studies have postulated that the cumulative stellar mass from these primordial building blocks, at $z\sim8$, is on the order of $\sim10^{6}-10^{7}$ M$_{\odot}$ (\citealp[][]{Mowla2024,Rusta2024}). Our estimate of $M_{*}\approx9\times10^{8}$ M$_{\odot}$ for the mass in proto-Milky Way populations (including the \textsl{Heracles} debris) is higher than these high-redshift measurements. We reason that the discrepancy between these two mass measurements is likely due to the redshift at which the \textsl{JWST} measurement were made ($z\sim8$) being much higher than the redshift predicted for a proto-Galaxy to form ($z\sim2$: \citealp[][]{Horta2024_proto}). This would lead to a smaller mass estimate than the one we determine in this work. Furthermore, the rapid growth in mass at early redshifts is consistent with a fast growth for the Milky Way at early times \citep[e.g.,][]{Mackereth2018}.

\section{Conclusions}
\label{sec:conclusions}
In this paper, we have set out to chemically identify stars belonging to the proto-Galaxy building blocks in order to then model their density and estimate a stellar mass measurement. Given our findings from Section~\ref{sec:chemistry} (Fig~\ref{fig:magnamal-fehs} through Fig~\ref{fig:rapo-mgfe}), but also recent theoretical results using cosmological simulations \citep{Horta2024_proto}, we define the proto-Galaxy to be the amalgamation of several building blocks: likely, the \textsl{Heracles} debris \citep[][]{Horta2021}; and the early progenitor of the Milky Way (\citealp[][]{Belokurov2022, Conroy2022}). All together, these comprise the poor-old heart \citep{Rix2022} of the Galaxy. \\

Our results can be summarised as follows:

\begin{itemize}
    \item We have found that by dissecting $\textsl{APOGEE-Gaia}$ RGB stars in the [Mg/Mn]-[Al/Fe] plane first (Fig~\ref{fig:magnamal-intro}), and then later in the [Mg/Fe]-[Fe/H] plane (Fig~\ref{fig:Tinsley}), we have been able to sub-select stars that predominantly belong to proto-Milky Way populations. 
    \item By examining the dependence of metallicity for metal-poor stars in the [Mg/Mn]-[Al/Fe] plane, we have noticed two striking features: $i$) stars in the high-$\alpha$/evolved region appear at metallicities as metal-poor as [Fe/H] $\sim-1.6$. These stars are the metal-poor extension of the high-$\alpha$ (\textit{in situ}) sequence, whose population formed rapidly with high star formation efficiency, thus reaching high [Al/Fe] whilst still being considerably metal-poor (roughly along the lines of the chemical evolution models presented in \citealt{Horta2021}). This population is present both in the inner Galaxy and in the Solar neighbourhood, and is likely associated with the main progenitor system of the Milky Way (also referred to as \textsl{Aurora}: \citealp[][]{Belokurov2022}). $ii$) there is a surplus of stars in the unevolved region of the [Mg/Mn]-[Al/Fe] plane at metallicities between $-1.6 < $[Fe/H]$ < -0.7$, whose stellar populations likely underwent slower chemical evolution. In the Solar neighbourhood, chemically unevolved stars with [Fe/H]$>-1.6$ likely correspond to the debris of the \textsl{Gaia-Enceladus/Sausage} accretion event, whereas in the inner Galaxy they are likely comprised by the \textsl{Heracles} system \citep{Horta2021}.
    
    \item Proto-Galaxy populations (including the \textsl{Heracles} debris), are well modelled by an oblate ($q\sim0.6$, $p\sim0.8$) Plummer model (Fig~\ref{fig:best-fits}), with a Plummer (scale-length) radius on the order of $\sim3.5$ kpc (Table~\ref{tab:fit-values}). This result supports the notion of a ``bulge'' component in the central Galaxy, if defined as a metal-poor, old, and centrally concentrated ellipsoid.
    \item We integrate the density within $r<10$ kpc for stars within $-2 < $ [Fe/H] $<-0.5$ to obtain a total minimum mass for proto-Galaxy populations of $M_{*}=9.1\pm0.2\times10^{8}$ M$_{\odot}$; for the main body of this population that sits between with -1.5 < [Fe/H] <-1, we estimate a mass of $M_{*}=7.2\pm0.2\times10^{8}$ M$_{\odot}$, that primarily comes from the Heracles debris. This implies that Heracles is a major building block of the Milky Way. However, it is also likely that we are not probing a fraction of the main progenitor system of the proto-Milky Way with the \textsl{APOGEE-Gaia} data.
\end{itemize} 

The advent of more precision abundance data for metal-poor stars in the inner $\sim5$ kpc of the Galaxy will enable further characterisation of the properties, and components, of the proto-Galaxy. The fifth phase of the Sloan Digital Sky Survey (\textsl{SDSS}: \citealp[][]{Kollmeier2017}) will be perfectly positioned for this task, as it is predicted to deliver precision chemistry for over $\sim70,000$ stars with [Fe/H] $<-1$. In addition, surveys such as \textsl{4MOST} \citep{DeJong2019}, \textsl{MOONS} \citep{moons}, and now \textsl{Gaia} \citep{Gaia2022} will also supply useful spectroscopic data for metal-poor stars in these inner regions.

\section*{Acknowledgements}
The authors would like to warmly thank the anonymous reviewer for providing and insightful and constructive referee report. DH thanks James Lane for help with obtaining the necessary isochrones to perform the density modelling fits, Julie Imig with help on running the DR17 version of the \texttt{apogee} code, and Maddie Lucey and Sergey Koposov for helpful discussions. He also deeply thanks Sue, Alex, and Debra for their constant support. The Flatiron Institute is funded by the Simons Foundation.

\appendix

\section{Density fits to chemically unevolved stars}
\label{appen_fits}

The density profiles tested in this work can be expressed analytically in the following manner: the cored power law profile can be written as

\begin{equation}
    \nu_{*}(r) \propto \Big( 1 + \frac{r}{r_{\mathrm{s}}}\Big)^{-\alpha},
\end{equation}

where $r_{\mathrm{s}}$ is the scale-length and $\alpha$ is the slope. The Einasto profile is described as 

\begin{equation}
    \nu_{*}(r) \propto \exp\Bigg[-d_{n} \Bigg(\Big(\frac{r}{r_{s}}\Big)^{1/n} - 1\Bigg)\Bigg]
\end{equation}
where $d_{n} = 3n - 0.3333 + 0.0079/n$ for $n>0.5$ \citep{Graham2006}, and the slope of the Einasto profile can be defined as 

\begin{equation}
    \alpha_{\mathrm{Ein}} = -\frac{d_{n}}{n}\Big(\frac{r}{r_{s}}\Big)^{1/n}.
\end{equation}

The exponential disc profile is expressed as

\begin{equation}
    \Sigma_{*}(R,z) \propto \exp \Bigg[ - \Bigg(\frac{(R - R_{0})}{h_{R}} + \frac{|z - z_{0}|}{h_{z}} \Bigg) \Bigg]
\end{equation}

where $h_{R}$ and $h_{z}$ are the scale-length and scale-height parameters, respectively. 

The Plummer profile can be written as

\begin{equation}
     \nu_{*}(r) \propto \frac{3}{4\pi a^{3}}\Bigg(1 + \frac{r^{2}}{a^{2}}\Bigg)^{-5/2}
\end{equation}
where $a$ is the Plummer radius. Similarly, the Hernquist profile can be expressed as

\begin{equation}
     \nu_{*}(r) \propto \frac{1}{2\pi}\frac{a}{r(r+a)^{3}}
\end{equation}

where a is the scale radius. Lastly, the S\'ersic profile is described as

\begin{equation}
    \nu_{*}(r) \propto \exp \Bigg[-b_{n}\Bigg(\Big(\frac{r}{r_{\mathrm{eff}}}\Big)^{1/n} -1 \Bigg)\Bigg]
\end{equation}
where $n$ is the S\'ersic index, $r_{\mathrm{eff}}$ is the effective radius, and $b_{n}$ is a constant such that $r_{\mathrm{eff}}$ is set to the half-light radius.

Furthermore, in the recent studies \citet{Mackereth2020} and \citet{Lane2023}, the density of \textsl{APOGEE} red giant branch stars, and particularly those on highly eccentric ($e>0.7$) orbits (i.e., the GES debris), were shown to be well modelled by a single triaxial and tilted power law. This model form is fully expressed as:

\begin{equation}
\label{eq:halo_model}
    \nu_{*}(r_{e}) \propto (1 - f_{\mathrm{disc}}) r_{e}^{-\alpha} \exp{(-\beta~r_{e})} + f_{\mathrm{disc}}\nu_{*,\mathrm{disc}},
\end{equation}

where $r_{e} = X + \frac{Y}{p} + \frac{Z}{q}$, $\beta$ is a cut-off scale parameter, and $f_{\mathrm{disc}}$ is a fraction parameter that informs how much contamination is accounted for in the data by an exponential disc profile with $h_{z} = 0.8$ kpc and $h_{R}=2.2$ kpc. 

\begin{figure*}
\centering
    \includegraphics[width=\textwidth]{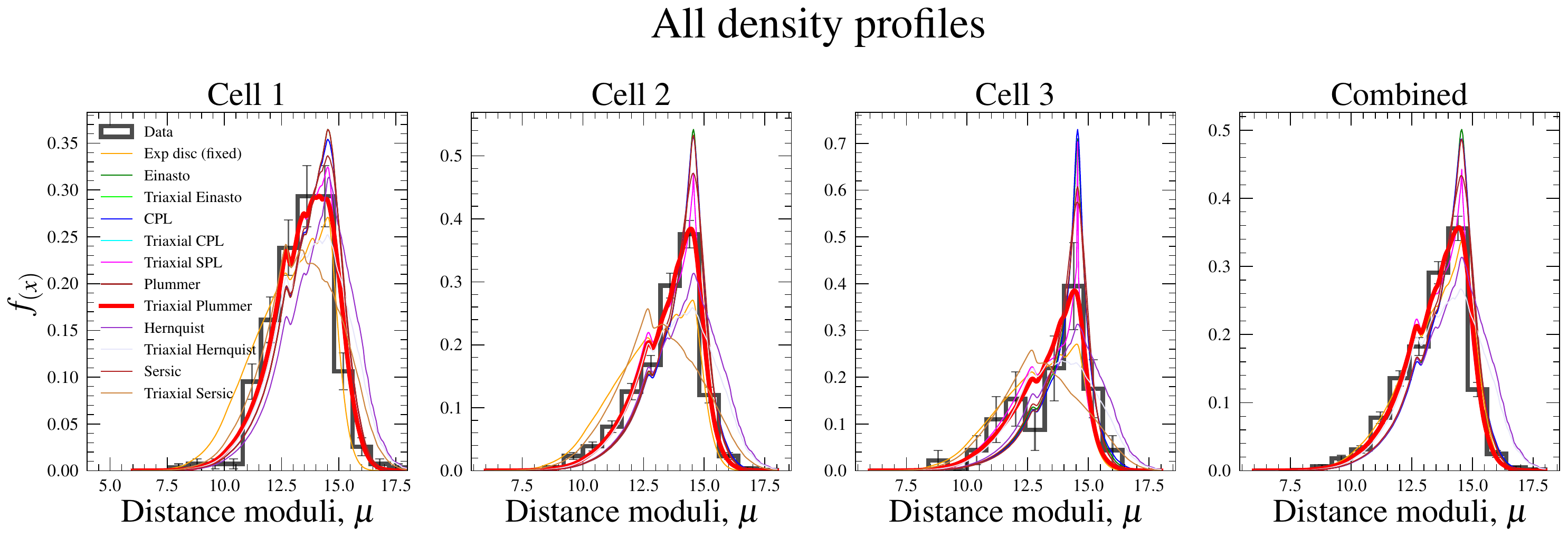}
\caption{Same as Fig~\ref{fig:best-fits}, but now for all models tested. The best fitting model (Triaxial Plummer) is shown as a thicker (red) line.}
    \label{fig:best-fits-all} 
\end{figure*}

\begin{figure*}
\centering
    \includegraphics[width=0.9\textwidth]{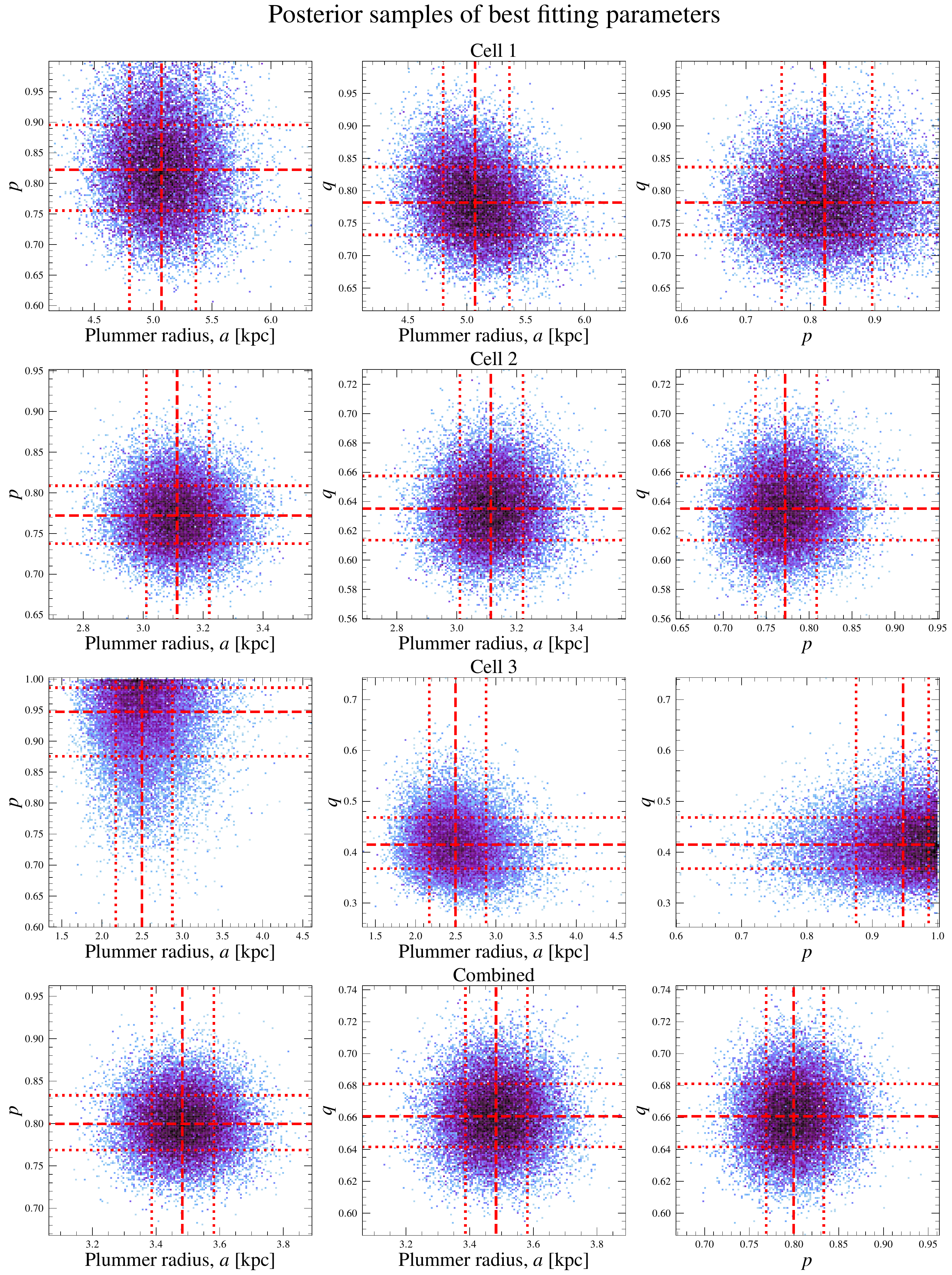}
\caption{2D density projection for the posterior samples of the parameters in the best fitting model (triaxial Plummer) for cells 1-3. The median(16,84) percentiles are shown as the dashed(dotted) lines. The posterior samples are well behaved, and converge to a given parameter value (see Table~\ref{tab:fit-values}).}
    \label{fig:params-best-fits} 
\end{figure*}

Fig~\ref{fig:best-fits-all} shows the data compared to all the density models tested for chemical cells 1-3. As can be seen by eye, but also estimated using a combination of the maximum negative log-likelihood, BIC, and AIC (Table~\ref{tab:bic}), the triaxial Plummer model is the best fitting profile. Moreover, the posterior samples from the MCMC method are shown in Fig~\ref{fig:params-best-fits}. The samples are well behaved and converge to a given parameter value (Table~\ref{tab:fit-values}).

\begin{table}
	\centering
	\caption{From left to right: density profile, the negative maximum log-likelihood, the Bayesian Information criterion (BIC), and the Akaike Information Criterion (AIC). We report the values for cell 1/ cell 2/ cell 3, respectively. The bolded profile is the chosen best fitting profile based on a combination of the three metrics. Missing values mean that the model didn't converge.}
	\begin{tabular}{lcccccr} 
		\hline
		Density profile & $-$ln($\mathcal{L_{\mathrm{max}}}$) & BIC & AIC\\
    
   \hline
    CPL & 2921/7580/423 & 27/31/20& 20/22/16\\
    Tri-CPL &  --/--/--& --/--/--& --/--/--\\
    Einasto &  2921/7574/422& 27/31/20& 20/22/16\\
    Tri-Einasto & --/--/--& --/--/--& --/--/--\\
    Plummer & 2923/7613/429 & 21/25/16& 18/20/14\\
 \textbf{Tri-Plummer} & 2914/7529/409 & 33/38/24& 22/24/18\\
    Hernquist & 2994/7993/455 & 21/25/16& 18/20/14\\
    Tri-Hernquist & 2991/7973/451 & 33/38/24& 22/24/18\\
    S\'ersic & 2921/7574/423 & 33/38/24& 22/24/18\\
    Tri-S\'ersic & 3161/8698/494 & 45/52/33& 26/28/22\\
    Tri-SPL w./ disc + cutoff & 2907/7485/411 & 62/72/44& 32/34/28\\
\hline
	\end{tabular}
 \label{tab:bic}
\end{table}

\begin{table*}
	\centering
	\caption{Density profile and optimised parameters for all models tested in this work. We report the values for cell 1, cell 2, and cell 3, respectively. Values with -- correspond to not optimised fits.}
	\begin{tabular}{lcccccr} 
		\hline
		Density profile & Cell 1 & Cell 2 & Cell 3\\
    
   \hline
    CPL & $\alpha = 8.9; r_{s} = 12.6$ (kpc) & $\alpha = 5.1; r_{s} = 2.8$ (kpc) & $\alpha = 4.0; r_{s} = 0.9$ (kpc)\\
    Tri-CPL & -- & -- & --\\
    Einasto &  $n = 1.6; r_{s} = 6.0$ (kpc)& $n = 2.5; r_{s} = 3.8$ (kpc)& $n = 3.5; r_{s} = 2.8$ (kpc)\\
    Tri-Einasto & -- & -- & --\\
    Plummer & $a = 4.8$ (kpc) & $a = 3.2$ (kpc) & $a = 2.6$ (kpc)  \\
 \textbf{Tri-Plummer} & $a=5.1$ (kpc); $p=0.8$; $q=0.8$ & $a=3.1$ (kpc); $p=0.8$; $q=0.6$ & $a=2.5$ (kpc); $p=0.9$; $q=0.4$ \\
    Hernquist & $a = 0.1$ (kpc) & $a = 0.1$ (kpc) & $a = 0.1$ (kpc)\\
    Tri-Hernquist  & $a=0.1$ (kpc); $p=1$; $q=0.5$ & $a=0.1$ (kpc); $p=1$; $q=0.5$ & $a=0.1$ (kpc); $p=1$; $q=0.4$\\
    S\'ersic & $n = 1.6; b_{n} = 1.4; r_{\mathrm{eff}} = 0.9$ (kpc) & $n = 1.4; b_{n} = 1.5; r_{\mathrm{eff}} = 0.1$ (kpc) & $n = 2.2; b_{n} = 1.3; r_{\mathrm{eff}} = 0.1$ (kpc)\\
    Tri-S\'ersic & $n = 1.2; b_{n} = 0.0; r_{\mathrm{eff}} = 7.2$ (kpc)& $n = 1.4; b_{n} = 0.0; r_{\mathrm{eff}} = 10.0$ (kpc)& $n = 1.0; b_{n} = 0.0; r_{\mathrm{eff}} = 5.3$ (kpc)\\
    & $p = 0.0; q = 0.0$ & $p = 0.0; q = 0.0$ & $p = 0.0; q = 0.0$ \\
    Tri-SPL w./ disc + cutoff & $\alpha = 1.5, \beta = 0.2, p = 0.9, q = 0.7$ & $\alpha = 2.1, \beta = 0.2, p = 0.8, q = 0.6$ & $\alpha = 2.7, \beta = 0.1, p = 0.8, q = 0.4$\\ 
    & $\theta = 0.0 (^{\circ}), \phi = 0.3 (^{\circ}), \eta = 0.1$ & $\theta = 0.4 (^{\circ}), \phi = 0.3 (^{\circ}), \eta = 0.0$ & $\theta = 0.0 (^{\circ}), \phi = 0.0 (^{\circ}), \eta = 0.0$ \\
    & $f_{\mathrm{disc}} = 0.0 (\%)$ & $f_{\mathrm{disc}} = 0.0 (\%)$ & $f_{\mathrm{disc}} = 0.0 (\%)$\\
    
\hline
	\end{tabular}
 \label{tab:bic}
\end{table*}

\begin{table*}
	\centering
	\caption{Density profile and optimised parameters for all models tested in this work, but now for the combined sample. Values with -- correspond to not optimised fits.}
	\begin{tabular}{lcccccr} 
		\hline
		Density profile & Combined\\
    
   \hline
    CPL & $\alpha = 5.2; r_{s} = 3.2$ (kpc) \\
    Tri-CPL & -- \\
    Einasto &  $n = 2.5; r_{s} = 4.2$ (kpc)\\
    Tri-Einasto & -- \\
    Plummer & $a = 3.5$ (kpc) \\
 \textbf{Tri-Plummer} & $a=3.5$ (kpc); $p=0.8$; $q=0.7$\\
    Hernquist & $a = 0.1$ (kpc) \\
    Tri-Hernquist  & $a=0.1$ (kpc); $p=1$; $q=0.6$ \\
    S\'ersic & $n = 2.2; b_{n} = 1.2; r_{\mathrm{eff}} = 0.1$ (kpc) \\
    Tri-S\'ersic & -- \\
    Tri-SPL w./ disc + cutoff & $\alpha = 2.1, \beta = 0.2, p = 0.9, q = 0.6$ \\ 
    & $\theta = 0.0 (^{\circ}), \phi = 0.2 (^{\circ}), \eta = 0.1$ \\
    & $f_{\mathrm{disc}} = 0.0 (\%)$ \\
    
\hline
	\end{tabular}
 \label{tab:bic}
\end{table*}

\section*{Data Availability}
All \textsl{APOGEE} DR17 data used in this study is publicly available and can be found at: https: /www.sdss.org/dr17/. All \textsl{Gaia} DR3 data can be found at https://www.cosmos.esa.int/web/gaia/dr3.

\vspace{2mm}
\textit{Facilities:} \textsl{SDSS-IV} \citep{Blanton2017} \textsl{Apache Point Observatory} \citep{Gunn2006}, \textsl{Las Campanas Observatory} \citep{BowenVaughan1973}, \textsl{Gaia} \citep{Gaia2018,Gaia2020,Gaiaedr3}

\textit{Software}:
    \texttt{matplotlib} \citep{Hunter:2007},
    \texttt{numpy} \citep{NumPy},
    \texttt{scipy} \citep{SciPy},
    \texttt{gala} \citep{Price2017},
    \texttt{apogee} \citep{Bovy2016}.
    \texttt{}



\bibliographystyle{mnras}
\bibliography{refs}

\appendix

\label{lastpage}
\end{document}